\documentclass[sigconf]{acmart}
\usepackage[utf8]{inputenc}
\usepackage[english]{babel}
\usepackage[T1]{fontenc}
\usepackage{graphicx}
\usepackage{subcaption}
\usepackage{tikz}
\usepackage{pgfplots}
\usepackage{booktabs}
\def\BibTeX{{\rm B\kern-.05em{\sc i\kern-.025em b}\kern-.08emT\kern-.1667em\lower.7ex\hbox{E}\kern-.125emX}}
    
\copyrightyear{2019}
\acmYear{2019}
\setcopyright{acmlicensed}
\acmConference[ARES '19]{ARES '19: Proceedings of the 14th International Conference on Availability, Reliability and Security }{August 26--29, 2019}{Canterbury, NY}
\acmBooktitle{ARES '19: Proceedings of the 14th International Conference on Availability, Reliability and Security, August 26--29, 2019, Canterbury, NY}

\begin{document}

%
\title{Detecting DGA domains with recurrent neural networks and side
information}

%
\author{Ryan R. Curtin}
\affiliation{
  \institution{Center for Advanced Machine Learning\\Symantec Corporation}
  \streetaddress{1200 Abernathy Road, Ste. 1700}
  \city{Atlanta}
  \state{Georgia}
  \country{USA}
  \postcode{30328-5671}
}

\author{Andrew B. Gardner}
\affiliation{
  \institution{Center for Advanced Machine Learning\\Symantec Corporation}
  \streetaddress{1200 Abernathy Road, Ste. 1700}
  \city{Atlanta}
  \state{Georgia}
  \country{USA}
  \postcode{30328-5671}
}

\author{Slawomir Grzonkowski}
\affiliation{
  \institution{Targeted Attack Analytics\\Symantec Corporation}
  \streetaddress{Ballycoolin Business Park}
  \city{Dublin}
  \country{Ireland}
  \postcode{15}
}

\author{Alexey Kleymenov}
\affiliation{
  \institution{Targeted Attack Analytics\\Symantec Corporation}
  \streetaddress{Ballycoolin Business Park}
  \city{Dublin}
  \country{Ireland}
  \postcode{15}
}

\author{Alejandro Mosquera}
\affiliation{
  \institution{Targeted Attack Analytics\\Symantec Corporation}
  \streetaddress{Reading International Business Park}
  \city{Reading}
  \state{Berkshire}
  \country{UK}
  \postcode{RG2 6DA}
}

%
\renewcommand{\shortauthors}{R.R. Curtin et~al.}

%
\begin{abstract}
Modern malware typically makes use of a domain generation algorithm (DGA) to
avoid command and control domains or IPs being seized or sinkholed.  This means
that an infected system may attempt to access many domains in an attempt to
contact the command and control server.  Therefore, the automatic detection of
DGA domains is an important task, both for the sake of blocking malicious
domains and identifying compromised hosts.  However, many DGAs use English
wordlists to generate plausibly clean-looking domain names; this makes automatic
detection difficult.  In this work, we devise a notion of difficulty for DGA
families called the {\it smashword score}; this measures how much a DGA family
looks like English words.  We find that this measure accurately reflects how
much a DGA family's domains look like they are made from natural English words.
We then describe our new modeling approach, which is a combination of a novel
recurrent neural network architecture with domain registration side information.
Our experiments show the model is capable of effectively identifying domains
generated by difficult DGA families.  Our experiments also show that our model
outperforms existing approaches, and is able to reliably detect difficult DGA
families such as {\tt matsnu}, {\tt suppobox}, {\tt rovnix}, and others.  The
model's performance compared to the state of the art is best for DGA families
that resemble English words.  We believe that this model could either be used in
a standalone DGA domain detector---such as an endpoint security application---or
alternately the model could be used as a part of a larger malware detection
system.

\end{abstract}

%
%


%

%
\maketitle

\section{Introduction}


\begin{figure}[t!]
\centering
\includegraphics[width=0.4\textwidth]{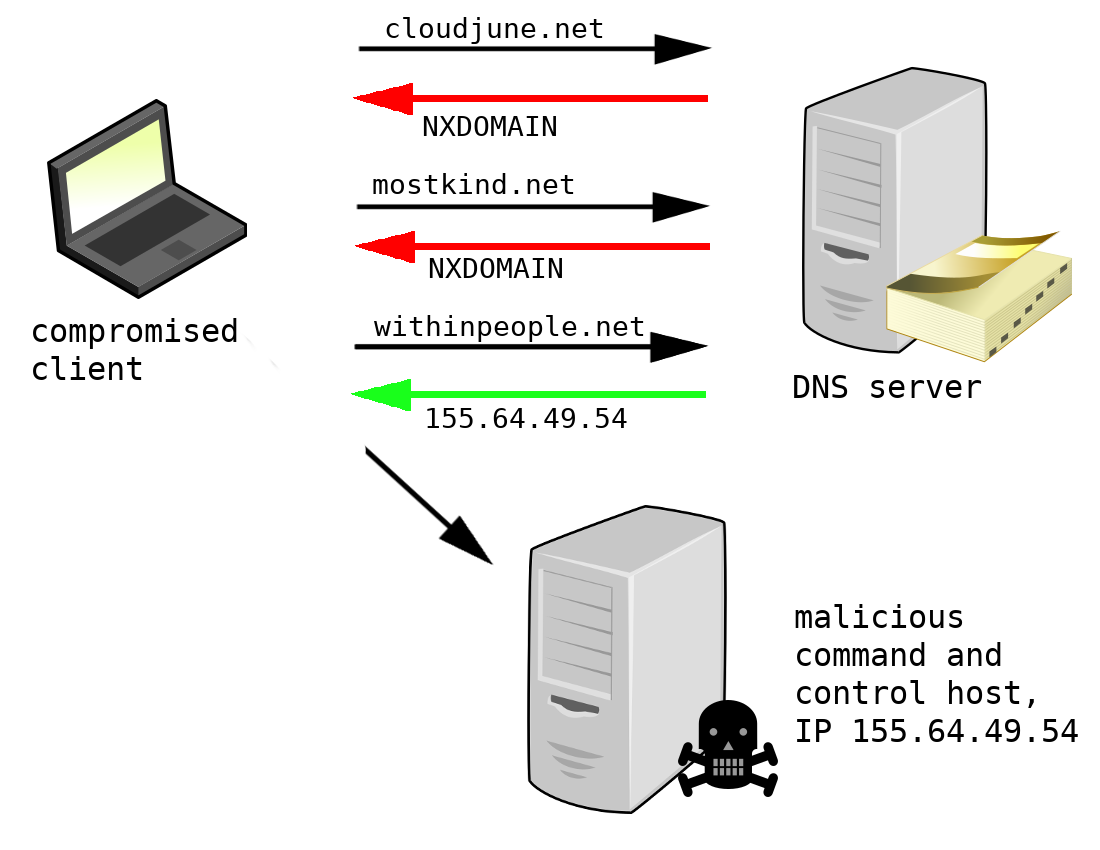}
\label{fig:dga-process}
\caption{Typical example of malware using a DGA to find its command-and-control
(C\&C) server.  The infected host attempts to resolve a number of DGA-generated
domains, and connects to the first one that resolves successfully.}
\end{figure}

Many modern malware families communicate with a centralized command and control
(C\&C) server.  In order to do this, the malware must know the location of the
C\&C server to connect to---simple approaches might hardcode an IP or a domain
name.  But, these are easy to mitigate: the traffic to a specific IP can be
trivially blocked, and domain names can be easily seized.  Therefore, modern
malware authors use {\it domain generation algorithms} (DGAs) in order to
generate a large set of possible domain names where the C\&C server may exist.

Typically, an infected machine will use the DGA to serially generate domain
names.  Each of these domain names will be resolved, and if the DNS resolution
does not result in an NXDOMAIN response (i.e. if the domain name is registered),
then the machine will attempt to connect to the resolved IP as if it is the C\&C
server.  If any step of that process is not successful, then the machine will
generate another domain name with the DGA and try again, until it is successful.
Some DGA families generate random-looking domain names such as {\tt
xobxceagb[.]biz}; others generate difficult-to-distinguish domain names like
{\tt dutykind[.]net}.

This DGA-based approach to finding the C\&C server is robust to IP blocking and
domain name seizure; the C\&C server operator can use any IP they have access to
(and they may use different IPs at different times), and typically the number of
unique domain names a DGA can generate is quite large, and sometimes the DGA
itself may be hard to reverse engineer.  Therefore, it is not generally feasible
to pre-emptively seize all domain names that a DGA could generate.  In fact,
DGAs may even generate domain names that are not malicious or compromised, and
this does not affect the malware's ability to reach the C\&C server eventually.

As a consequence, the task of determining whether or not a given domain name is
produced by a DGA is an integral part of modern malware defenses.
Simultaneously, as DGA authors create DGAs that generate domain names that do not 
look randomly-generated, the challenge of detecting these domain names increases.

A large body of related work seeks to use machine learning techniques directly
to classify domains as generated by DGAs or not.  Our contribution adds to this
lineage of work; here, our machine learning detector is one component of an
effective malware detection system.  To this end, we describe a machine learning
system that is able to
accurately classify a domain name as DGA-generated or clean using only the
domain name itself and some simple additional features derived from WHOIS data.
This system is especially effective on DGA families that generate domain names
based on English word lists (i.e., domains that look benign to a human observer).
Compared to previous approaches, our system performs better on
difficult-to-detect DGA families that resemble English words (such as the {\tt
matsnu} and {\tt suppobox} families), and the system is not difficult to deploy
in a real-world environment---either as a standalone detector or as part of a
larger malware detection system.

Overall, this paper makes the following contributions:
\begin{itemize}
\itemsep 5pt
\item{We provide a novel machine learning system built partially on recurrent
neural networks that is capable of classifying DGA-generated domain names even
from families traditionally understood as difficult.  To achieve this degree of
performance, our model takes advantage of side information such as WHOIS.}
\item{This model is robust: although it is trained with WHOIS information,
predictions can still be made if WHOIS or other network level information is not
available.  This is crucial for real-time detection and prevention of malware
outbreaks.}
\item{We devise a new measure that we term the {\it smashword score}.  We rank
41 DGAs in terms of detection difficulty using this measure, giving an intuitive
measure of difficulty related to how closely the domain resembles English words.
Our approach can be re-used for new DGA families, and we believe our measure is
useful for other DGA detection works in the future.}
\item{We successfully classify difficult DGA-generated domains using our model
that other state-of-the-art approaches could not conclusively label; this
includes domains with high smashword scores (e.g., those that are composed of
combinations of English words).  Note that these domains can even be difficult
for humans to classify correctly.}
\end{itemize}

\section{Related Work}

The problem of distinguishing legitimate domain names from algorithmically generated is certainly not new, and has
been studied for a number of years.  DGAs first became widely known to the
community with the introduction of Kraken \cite{kraken09} and Conficker \cite{conficker2009}
in 2008.  Since that time, DGAs in malware have proliferated.

The early efforts to stop this threat were dealing with lack of sufficient training data to apply machine learning approaches~\cite{avivbotnet11}. One decade later it continues to be a problem but to a smaller extent.
Thus early proposed approaches and techniques were rather statistical. For example Yadav et al.~\cite{domainflux12}
applied such technique to show differences between valid domain names and algorithmically generated ones. The limitation of such approach would be that it often does not transfer to a different DGA family.

Another milestone in detection techniques was credited to more extensive usage of DNS data. For example Zhou et al.~\cite{Zhou2013DGABasedBD} gathered DNS NXDOMAIN data from RDNSs and then used it to assemble a set of suspicious algorithmically generated domain names.

A different approach was proposed by Jian et al.~\cite{dnsgraph10}. It relied on DNS traffic analysis but only for failed lookups. In this technique interactions between hosts and failed domain names would be extracted. Then a graph decomposition algorithm using nonnegative matrix tri-factorization technique to iteratively extract coherent co-clusters would be applied. The obtained sub-graphs would be further analyzed by exploring temporal properties of generated clusters. The authors claim that their anomaly based approach can detect new and previously undiscovered threats.

Further research efforts evolved towards more and more extensive usage of machine learning techniques. At a large scale, it was pioneered by Phoenix~\cite{phoenix14} that was able to use both the URLs and other side information to detect DGA botnets. The list of parameters observed by this system includes some handcrafted features like pronounceability, blacklist information, DNS query information. This approach does not use any recurrent neural network (RNN) or powerful modeling technique for the domains themselves leaving a room for improvement. Tong and Nguyen \cite{semanticdga16} have already proposed extensions to the Phoenix system. They included additional measures such as entropy, n-grams and modified distance metric for domain classification.

Further progress in DGA detection was reported when using machine learning techniques. For example, Zhao et al.~\cite{aptdns15} addressed the problem in the context of detecting APT malware.
The authors proposed 14 features based on their big data research to characterize different properties of malware-related DNS and the ways that they are queried as well as defined network traffic features that can identify the traffic of compromised clients that have remotely been controlled.
The features are comprised of signature-based engine, anomaly-based engine and
so-called dynamic DNS features. The data was filtered by using
Alexa\footnote{See \url{https://www.alexa.com/topsites}.} popularity and prevalence based on the number of hosts connecting to domains.
As the outcome, an engine was built and it was used to compute reputation scores for IP addresses using extracted features vectors.
The results are produced by using the J48 decision tree algorithm.

A comparable approach was presented by Luo et al.~\cite{dgasensor17} who described a system using lexical patterns that were extracted from clean domains listed in the Alexa top 100k domains as well as confirmed malicious DGA cases. The proposed approach is machine learning-based and achieves 93\% accuracy for detecting malicious domains on the test dataset.

Additional improvements for state of the art results were reported by Woodbridge et al.~\cite{woodbridge2016predicting}.
Despite a relatively simple Long Short-Term Memory (LSTM) network used to
classify DGA domains, the approach was reported to have a high level of
effectiveness. The presented results still have certain shortcomings, especially
for difficult DGA classes that resemble English words.

The same problem was approached from a different angle by Anderson et al.~\cite{deepdga16}.
The authors use a Generative Adversarial Network (GAN) to generate adversarial DGA domain names to try and deceive a classifier. The authors were able to achieve this goal. Then the GAN-generated domain names were added to the training set, which resulted in improved DGA detection performance.
However, the authors did not test on any DGA families that look like they are
made up of English words.

Shibahara et al.~\cite{Shibahara2016EfficientDM} proposed a slightly different
algorithm that is using RNN on changes in network communication with a goal of
reducing malware analysis time. This approach is not DGA-only specific but
rather generic and attempts to cover other types of malware. However, it could
successfully be used against DGA-type of threats based on their communication
patterns. Thus this technique requires additional run-time data that is
not required in many of the other approaches as it requires malware sandboxing.
The authors claim that without their optimization the analysis time takes over
15 min. and their approach reduces this time by 67\%, preserving the detection rate of malicious URLs at 97.9\%.

Overall, though the task of DGA detection is certainly not new, there has not
been much focus on directly detecting DGA families made from English words
using the domain itself as a feature. This task has been described as
`extremely difficult' in some previous works \cite{woodbridge2016predicting}.
Here, our focus is specifically on those DGA families.

\section{Measuring the difficulty of detection of a DGA family}


Since data-based approaches for the detection of malicious domains have been a
recurrent trend during recent years, it is inevitable that malware authors would
shift to generation algorithms that overlap with lexical patterns commonly found
in clean datasets to avoid being detected. Taking into account this adversarial
environment, we need to be able to measure how our DGA detection models will
perform not only overall, but also against the most difficult samples.
In this context, `difficult' samples can be understood to be those that trick
existing detectors---the most relevant example is those DGA families that combine
English words, like the {\tt matsnu} family \cite{skuratovich2015}, which was
one of the first of many families to build domain names from English word lists.
These generate domains like the natural-looking domains {\tt songneckspiritprintmetal[.]com} and {\tt
westassociatereplacerisk[.]com}, which present a much harder challenge to the many
detection systems that depend on lexical features~\cite{yu2018character,
woodbridge2016predicting, phoenix14, deepdga16}.

An exploratory data analysis of our dataset shows that DGA families have
characteristics that can affect the performance of classification
approaches. From an information theory point of view, both the average length
$\bar{l}(\cdot)$ and the average character entropy \cite{shannon48}
$\bar{c}(\cdot)$ of the domain names seem likely to be interesting features to
compare.  The entropy of a single domain $x$ is calculated as below:

\vspace*{-0.5em}
\begin{equation}
\hat{c}(x) = -\sum_{x_i \in x} p(x_i) \log_2 p(x_i)
\end{equation}
\vspace*{-0.5em}

\noindent where $p(x_i)$ is the empirical probability of the character $x_i$ in
the string $x$.  However, in our experiments, we found no serious correlation
between the average character entropy $\bar{c}(\cdot)$ of a DGA family and
if that family was made up of difficult English-like words.  Thus, we
cannot use $\bar{c}(\cdot)$ as a proxy for the difficulty of detecting a family.

Therefore, we have developed the {\it smashword score} $\hat{s}(\cdot)$, which
is the the average $n$-gram overlap (with $n$ ranging from 3-5) with words from an English dictionary.
%
%
The computation of the smashword score amounts to calculating term-frequency
inverse-document-frequency (TF-IDF) \cite{sparck1972statistical} scores for a
domain name using an English list of words as a reference document set.
Specifically:

\vspace*{-0.5em}
\begin{equation}
\hat{s}(x) = \frac{1}{| \mathcal{N}_{i,j}(x) |} \sum_{n_i \in
\mathcal{N}_{i,j}(x) \cap \mathcal{N}_{i,j}(D)} \log\left(|\{ d \in D : n_i \in d
\}|\right).
\end{equation}
\vspace*{-0.5em}

In this equation, $\mathcal{N}_{i,j}(x)$ refers to the set of character
$n$-grams in the domain $x$ of length $i$ or $j$, $D$ refers to the English
word list, and $\mathcal{N}_{i,j}(D)$ refers to the set of character $n$-grams in
the entire word list $D$ of length $i$ or $j$.  The $\log$ term is the count of
times an $n$-gram appears in the entire word list $D$.  If there is no overlap in any $n$-grams between the
domain and the word list, the score is has a lower bound of 0 and an upper bound depends on the word list.  The
score is normalized to the number of $n$-grams in the domain.

Computing the smashword score $\hat{s}(x)$ for a string $x$ can be done in
$\min(| \mathcal{N}_{3,4,5}(x) |, | \mathcal{N}_{3,4,5}(D) |)$ operations; but
since $| \mathcal{N}_{3,4,5}(x) |$ will generally be much smaller than $|
\mathcal{N}_{3,4,5}(D) |$, we can say that the computation will generally take
time linear in the length of the string $x$, since in a string with length
$|x|$, there are $|x| - 2$ 3-grams, $|x| - 3$ 4-grams, and $|x| - 4$ 5-grams.

The average smashword score $\bar{s}(\cdot)$ of a DGA family is then calculated
by simply taking the average smashword score $\hat{s}(\cdot)$ of all of the
domains in that family that are present in the data.


We can expect that domains with a high smashword score will resemble English
words, and thus we expect that $\bar{s}(\cdot)$ is a good indicator of the
difficulty of detecting a DGA domain.  Indeed, in the following section we find
that our data bears out this expectation.

\section{DGA Families}
\label{sec:families}

\begin{table*}[t!]
\begin{center}
\begin{tabular}{lccccll}
\toprule
{\bf DGA family} & {\bf $n$} & {\bf $\bar{l}(x)$} & {\bf $\bar{c}(x)$} & {\bf $\bar{s}(x)$} & {\bf sample 1} & {\bf sample 2} \\
\midrule
{\bf banjori}$^3$                      & 435385 & 22.165  & 3.767 & {\bf
173.909} & {\scriptsize {\tt iivnleasuredehydratorysagp[.]com}} & {\scriptsize {\tt yanzvinskycattederifg[.]com}} \\
beebone$^3$       & 210  & 13.223        & 3.466 &  66.159 & {\scriptsize {\tt backdates10[.]com}} & {\scriptsize {\tt dnsfor3[.]net}} \\
chinad$^6$                     & 256  & 19.843        & 3.882 &  19.565 & {\scriptsize {\tt evybt5gtf2tprvbi[.]info}} & {\scriptsize {\tt m5j42r6uiqov2dgm[.]biz}} \\
conficker$^5$     & 100000 & 11.754    & 3.208 &  19.988 & {\scriptsize {\tt jemmmpo[.]ws}} & {\scriptsize {\tt xobxceagb[.]biz}} \\
{\bf gozi}$^3$                       & 1212  & 23.262       & 3.550 & {\bf
222.240} & {\scriptsize {\tt questionibus[.]com}} & {\scriptsize {\tt chrisredemptisviros[.]com}} \\
locky$^5$                     & 8 & 14.125           & 3.226 &  20.800 & {\scriptsize {\tt pccibcjncnhjn[.]yt}} & {\scriptsize {\tt qtysmobytagnrv[.]it}} \\
{\bf matsnu}$^6$                    & 99995 & 30.527       & 3.757 & {\bf
332.581} & {\scriptsize {\tt dutytillboxpossessprogress[.]com}} & {\scriptsize {\tt dropbridgeexplorecraftgive[.]com}} \\
murofet$^6$                     & 102020 & 41.619      & 4.488 &  38.683 & {\scriptsize {\tt hyernzfvd10k57lyozotazkrp52gzfyp22eu[.]org}} & {\scriptsize {\tt m19e41hydyfxgtcxjyn10nynukulxdvhub18[.]com}} \\
necurs$^6$                    & 2048 & 17.029        & 3.541 &  31.256 & {\scriptsize {\tt gmwyfuhwqveqcbasvtj[.]in}} & {\scriptsize {\tt pysetkbwbryxbegmwg[.]eu}} \\
newgoz$^6$        & 1000 & 29.885        & 4.214 &  23.476 & {\scriptsize {\tt afkv141b7q87du27t2i1b91gfp[.]biz}} & {\scriptsize {\tt krx32h1jjusuatm2aetjkn6jn[.]org}} \\
others\_dga\_b$^7$    & 2775 & 19.233        & 3.485 &  90.369 & {\scriptsize {\tt qktpxl[.]info}} & {\scriptsize {\tt meatopen[.]net}} \\
proslikefan$^6$              & 130  & 11.584        & 3.225 &  20.768 & {\scriptsize {\tt rxxeqcoy[.]cc}} & {\scriptsize {\tt avhpdzz[.]com}} \\
pykspa$^6$                     & 5010  & 14.470       & 3.375 &  40.551 & {\scriptsize {\tt tdxiogn[.]org}} & {\scriptsize {\tt dwaejwfox[.]cc}} \\
qakbot$^6$                    & 5000 & 20.719        & 3.740 &  43.164 & {\scriptsize {\tt hfbtlwqlqvoywaknjksaaeeus[.]net}} & {\scriptsize {\tt wlwcrapplotshymcia[.]org}} \\
ramdo$^6$        & 100000 & 20.000      & 3.393 &  51.347 & {\scriptsize {\tt aaooekcoyysuouaa[.]org}} & {\scriptsize {\tt skmaogeyiwqgeyym[.]org}} \\
ramnit$^6$                    & 100  & 17.340        & 3.585 &  35.981 & {\scriptsize {\tt ckyioylutybvcxv[.]com}} & {\scriptsize {\tt ibvtknxochoyjidm[.]com}} \\
ranbyus$^6$                   & 80 & 18.750          & 3.684 &  35.726 & {\scriptsize {\tt cyedjumagsrrav[.]cc}} & {\scriptsize {\tt jxbdxeyxttdmcjagi[.]me}} \\
{\bf rovnix}$^6$       & 99764  & 26.797      & 3.685 & {\bf 284.222} & {\scriptsize {\tt coloniesgovernmentsthe[.]com}} & {\scriptsize {\tt tohavetheontheofassent[.]com}} \\
shiotob$^6$                   & 2001 & 16.566        & 3.655 &  20.998 & {\scriptsize {\tt cwitdw951w1n9cm[.]net}} & {\scriptsize {\tt 3pttjmaw2g[.]com}} \\
{\bf suppobox}$^6$ & 258  & 17.298        & 3.341 & {\bf 152.536} & {\scriptsize {\tt bartholomewalbertson[.]net}} & {\scriptsize {\tt dutykind[.]net}} \\
tinba$^6$                     & 101001 & 16.006      & 3.457 &  31.592 & {\scriptsize {\tt fuvfpkpwgjqj[.]com}} & {\scriptsize {\tt hosvsbvbveee[.]com}} \\
volatile$^3$              & 352 & 19.000         & 3.745 & 112.362 & {\scriptsize {\tt hpyersgtdobfla[.]info}} & {\scriptsize {\tt ergtydobflashp[.]info}} \\
\bottomrule
\end{tabular}
\end{center}
\caption{Information about a representative set of DGA families.  We include the number of samples
$n$, average length $\bar{l}(\cdot)$, the average entropy $\bar{c}(\cdot)$, the
average smash-word score $\bar{s}(\cdot)$, and two examples of the
domains found in the family.  Families with high smash-word score are given in
{\bf bold}.}
\vspace*{-2.5em}
\label{tab:families}
\end{table*}


Before introducing our proposed classifier and experiments, we introduce our
dataset of DGA families and clean domains in order to perform some exploratory
analyses.  In this section we establish the Ground Truth (GT) datasets for both
confirmed DGA and non-DGA domains.  Each of our sources are taken from public
locations, making our dataset straightforward to reproduce.

\subsection{DGA Ground Truth Set}
The GT for DGA domains consists of domains generated using Python
implementations of real-world malware families using various seeds if necessary
as an input, as well as domains collected from the wild. In order to have
sufficiently diverse coverage, the following entities were selected in
order to represent many wide-spread patterns of DGA domains seen
in-the-wild in 2017/2018:
\begin{itemize}
	\item random-looking 2nd level domain names
	\item random-looking 3rd level domain names with generic 2nd level domain (usually dynamic DNS provider)
	\item domain names comprised of random words (generally English)
\end{itemize}
The last type of domains was of our particular interest as lexically they are
virtually indistinguishable from legal domains which means that some extra
techniques are required for detection.

In almost all cases DGAs are using some sort of input seeds in order to either
randomize the output and don't generated the same domains twice, or make it
unpredictable for researchers to avoid blocking or sinkholing. Here are some of
the popular seeds:
\begin{itemize}
	\item current date and/or time
	\item value embedded into a sample/group of samples by campaign (usually one DWORD)
	\item string(s) available online either on a malware authors or public server
	\item 3rd party public online document (for example, The US Declaration of Independence, the Apple license, etc)
\end{itemize}

In addition, some work has been done to make sure that there are diverse top level domains (TLDs) represented as malware authors tend to use only some particular ones which may introduce substantial skew to our dataset.
Overall, we have collected 41 DGA families. Information on each family is given
in Table \ref{tab:families}, including the average entropy $\bar{c}(\cdot)$ and
average smashword score $\bar{s}(\cdot)$. The families are collected from
multiple sources and denoted in the table: DGArchive\footnote{See
\url{https://dgarchive.caad.fkie.fraunhofer.de/}.},
an implementation for the {\tt locky} family found on
Github\footnote{See \url{https://github.com/sourcekris/locky}.},
Andrey Abakumov's DGA repository on Github\footnote{See
\url{https://github.com/andrewaeva/DGA}.}, and Johannes Bader's DGA
implementations\footnote{See \url{https://johannesbader.ch} and
\url{https://github.com/baderj/domain\_generation\_algorithms}.}.
Smaller or unknown families were grouped as {\tt others\_dga} and {\tt
others\_dga\_b} \footnote{Sinkholed domains collected from public WHOIS
registration information containing {\tt jgou.veia@gmail.com} as the contact
email.}. All of this data is publicly available; our set of DGA
domains is reproducible. 


\subsection{Non-DGA Ground Truth Set}
For non-DGA domains the GT is comprised of domains found in the Alexa top 1 million sites
and the OpenDNS public domain lists\footnote{These lists can be found at
\url{https://github.com/opendns/public-domain-lists}.}, giving 1.02M clean
domains for the clean GT set.
There were multiple major problems that had to be addressed:

\begin{enumerate}
\itemsep 5pt
	\item some prevalent DGA domains can manage to get into the list
of world top popular domains,
	\item 3rd level domains should be covered separately,
	\item some DGA domains are so short that they collided with the non-DGA
domains, and
	\item some DGA domains use combinations of English words, so the chances
that they collide with non-DGA domains are quite high.
\end{enumerate}

In the last two cases, malware authors have no problem when collisions take
place, as in
any case malware will be waiting for a valid response from the C\&C before
following up. Just the opposite, such cases can make the work of security
engineers more complicated as they cannot simply ban all domains generated by the
DGA---since some domains are known to be clean.  In our dataset, we only found
12 such domains that existed in both the non-DGA and DGA sets.  It presents
no modeling problem to leave these points in both sets.

\section{Side Information}
\label{sec:side}


DGA families with a high average smashword score $\hat{s}(\cdot)$ are very hard
to classify based on the domain names alone.  In fact, human analysts may even
have a difficult time differentiating---for instance, it is plausible at first
glance that {\tt darkhope.net} might be a personal website for a 1990s-era
teenaged computer enthusiast.  In reality, that domain name is generated by the
{\tt suppobox} DGA.  Thus, we cannot hope to build an effective classification
system using single domain names alone.

Therefore, we augment our domain names with {\it side information}, which we
collect from the WHOIS database \cite{rfc3912}.  Specifically, given a domain
name, we perform a WHOIS lookup, and extract the following numeric or Boolean
features:

\medskip
\begin{itemize}
  \item {\tt has\_registrarname}: Boolean, indicates whether a registrar
name is available.
  \medskip
  \item {\tt has\_contactemail}: Boolean, indicates if any contact email is
available.
  \medskip
  \item \texttt{days\_since\_created}, \texttt{days\_since\_updated}, \\
\texttt{days\_until\_expiration}: numeric, the number of days since the domain was
created, updated, or until expiration
  \medskip
  \item {\tt status\_length}: numeric, length of the ``status'' field
  \medskip
  \item {\tt has\_registrant\_info}, {\tt has\_admincontact\_info}, \\ {\tt
has\_billingcontact\_info}, {\tt has\_techcontact\_info}, \\ {\tt
has\_zonecontact\_info}: Boolean, indicates whether contact
information is available.
  \medskip
  \item {\tt has\_registrar\_iana\_id}: indicates whether a registrar IANA ID is
given.
\end{itemize}
\medskip

Note that for a non-registered domain name (NXDOMAIN), the boolean features will
all be {\tt false}, and the numeric features will all be taken as 0.

We do not perform any semantic analysis on the content of the WHOIS record;
instead, we focus on those features most likely to give us information relevant
to DGAs and C\&C servers: temporal information about the registration, and
whether the domain itself is registered.  The features we are using roughly
match the type of features used by Ma et~al. \cite{ma2009beyond}.

For our dataset, we used a snapshot of collected WHOIS data
with 245M records.
For our clean domain names, we matched 927k domains (91.7\%) to
WHOIS data, and for the DGA domains, we matched only 2.3k domains (0.18\%)
to WHOIS data.  This is expected, given that most DGA domains are never
registered.  In practice, either a snapshot of WHOIS data (potentially updated
nightly) or on-demand access of the WHOIS data could be used, depending on the
scalability needs of the deployment.

In our dataset, DGA families have an average of 3.5\% of their domains
matched to WHOIS data; with the {\tt ramnit} family matching the highest
percentage at 84\%, and the {\tt pandex} family matching the lowest nonzero
percentage at 0.008\% (only 7 out of 91758 domains registered).  19 families,
totaling 321k domains, have no domains matched to any WHOIS data.

Although having matching WHOIS data for a domain is strongly correlated with
whether or not the domain arises from a DGA, note that a detector built to
classify a domain as malicious simply if there is no WHOIS data would not be
very effective: with our data, it would achieve a true positive rate (TPR) of
96.5\%, but with an unacceptably high false positive rate (FPR) of 8.3\%.
Though WHOIS data gives us good information, it is not sufficient for
prediction.

\subsection{WHOIS and GDPR}

After the passing of the European privacy bill GDPR \cite{gdpr}, it is unclear
how WHOIS lookups will be affected \cite{icann-gdpr}.  At the time of our
experiments, WHOIS data was still publicly available.  However, if this is not
the case in the future, it would be easy to find alternatives.  Given that the
important features we extract depend more on the temporal registration
information than the contact details of the registrant, we could replace our
WHOIS features with DNS tracking systems like Active DNS
\cite{kountouras2016enabling} or the Alembic system \cite{lever2016domain}.

At the time of this writing, it is not clear what the long-term solution for
WHOIS data will be.  But, since WHOIS data is widely used for security
applications \cite{ma2009beyond, bilge2011exposure, canali2011prophiler}, it
seems unlikely that the types of features we are using for our system will
become unavailable.

\begin{figure}[b!]
  \centering
  {\scriptsize
  \begin{tikzpicture}[scale=0.4]
    \begin{scope}
  \pgfsetstrokecolor{black}
  \definecolor{strokecol}{rgb}{1.0,1.0,1.0};
  \pgfsetstrokecolor{strokecol}
  \definecolor{fillcol}{rgb}{1.0,1.0,1.0};
  \pgfsetfillcolor{fillcol}
  \filldraw (0.0bp,0.0bp) -- (0.0bp,281.0bp) -- (489.08bp,281.0bp) -- (489.08bp,0.0bp) -- cycle;
\end{scope}
\begin{scope}
  \pgfsetstrokecolor{black}
  \definecolor{strokecol}{rgb}{1.0,1.0,1.0};
  \pgfsetstrokecolor{strokecol}
  \definecolor{fillcol}{rgb}{1.0,1.0,1.0};
  \pgfsetfillcolor{fillcol}
  \filldraw (0.0bp,0.0bp) -- (0.0bp,281.0bp) -- (489.08bp,281.0bp) -- (489.08bp,0.0bp) -- cycle;
\end{scope}
  \pgfsetcolor{black}
  \draw [->] (185.28bp,158.86bp) .. controls (190.54bp,149.11bp) and (197.52bp,137.06bp)  .. (204.8bp,127.0bp) .. controls (207.29bp,123.56bp) and (210.05bp,120.07bp)  .. (219.53bp,109.06bp);
  \definecolor{strokecol}{rgb}{0.0,0.0,0.0};
  \pgfsetstrokecolor{strokecol}
  \draw (236.13bp,134.0bp) node {likelihoods};
  \draw [->] (282.99bp,244.92bp) .. controls (298.35bp,239.51bp) and (314.96bp,233.33bp)  .. (330.02bp,227.0bp) .. controls (349.53bp,218.81bp) and (370.6bp,208.7bp)  .. (397.55bp,195.14bp);
  \draw (420.69bp,220.0bp) node {\tt www.website.com};
  \draw [->] (288.0bp,158.58bp) .. controls (282.06bp,148.98bp) and (274.45bp,137.14bp)  .. (267.02bp,127.0bp) .. controls (264.65bp,123.77bp) and (262.09bp,120.45bp)  .. (253.05bp,109.33bp);
  \draw (313.64bp,134.0bp) node {one-hot TLD};
  \draw [->] (178.29bp,244.85bp) .. controls (164.99bp,239.53bp) and (150.77bp,233.41bp)  .. (137.97bp,227.0bp) .. controls (122.29bp,219.15bp) and (105.67bp,209.32bp)  .. (83.056bp,195.08bp);
  \draw (165.54bp,220.0bp) node {\tt www};
  \draw [->] (215.64bp,244.85bp) .. controls (211.88bp,239.19bp) and (207.79bp,232.88bp)  .. (204.2bp,227.0bp) .. controls (199.74bp,219.71bp) and (195.1bp,211.69bp)  .. (185.88bp,195.23bp);
  \draw (236.43bp,220.0bp) node {\tt website};
  \draw [->] (237.02bp,72.955bp) .. controls (237.02bp,64.883bp) and (237.02bp,55.176bp)  .. (237.02bp,36.09bp);
  \draw [->] (249.63bp,244.87bp) .. controls (255.58bp,239.41bp) and (261.81bp,233.22bp)  .. (267.02bp,227.0bp) .. controls (272.92bp,219.96bp) and (278.59bp,211.76bp)  .. (288.82bp,195.36bp);
  \draw (301.43bp,220.0bp) node {\tt com};
  \draw [->] (407.72bp,158.98bp) .. controls (392.81bp,148.55bp) and (372.99bp,135.8bp)  .. (354.02bp,127.0bp) .. controls (339.71bp,120.36bp) and (323.86bp,114.55bp)  .. (298.61bp,106.54bp);
  \draw (425.86bp,134.0bp) node {WHOIS features};
  \draw [->] (79.629bp,158.87bp) .. controls (93.882bp,148.54bp) and (112.73bp,135.91bp)  .. (130.8bp,127.0bp) .. controls (141.89bp,121.53bp) and (154.02bp,116.56bp)  .. (175.52bp,108.79bp);
  \draw (162.13bp,134.0bp) node {likelihoods};
\begin{scope}
  \definecolor{strokecol}{rgb}{0.0,0.0,0.0};
  \pgfsetstrokecolor{strokecol}
  \draw (222.11bp,195.0bp) -- (129.93bp,195.0bp) -- (129.93bp,159.0bp) -- (222.11bp,159.0bp) -- cycle;
  \draw (176.02bp,179.8bp) node {GLRT LSTM};
  \draw (176.02bp,165.8bp) node {(for domains)};
\end{scope}
\begin{scope}
  \definecolor{strokecol}{rgb}{0.0,0.0,0.0};
  \pgfsetstrokecolor{strokecol}
  \draw (489.14bp,195.0bp) -- (374.9bp,195.0bp) -- (374.9bp,159.0bp) -- (489.14bp,159.0bp) -- cycle;
  \draw (432.02bp,179.8bp) node {Matched WHOIS};
  \draw (432.02bp,165.8bp) node {side information};
\end{scope}
\begin{scope}
  \definecolor{strokecol}{rgb}{1.0,1.0,1.0};
  \pgfsetstrokecolor{strokecol}
  \draw (423.07bp,36.0bp) -- (50.97bp,36.0bp) -- (50.97bp,0.0bp) -- (423.07bp,0.0bp) -- cycle;
  \definecolor{strokecol}{rgb}{0.0,0.0,0.0};
  \pgfsetstrokecolor{strokecol}
  \draw (237.02bp,20.8bp) node {output classification};
  \draw (237.02bp,6.8bp) node {e.g. $P(\mathrm{clean}) = 0.99; P(\mathrm{malicious}) = 0.01$};
\end{scope}
\begin{scope}
  \definecolor{strokecol}{rgb}{0.0,0.0,0.0};
  \pgfsetstrokecolor{strokecol}
  \draw (357.21bp,195.0bp) -- (240.83bp,195.0bp) -- (240.83bp,159.0bp) -- (357.21bp,159.0bp) -- cycle;
  \draw (299.02bp,179.8bp) node {One-hot encoding};
  \draw (299.02bp,165.8bp) node {(for TLDs)};
\end{scope}
\begin{scope}
  \definecolor{strokecol}{rgb}{1.0,1.0,1.0};
  \pgfsetstrokecolor{strokecol}
  \draw (309.58bp,281.0bp) -- (146.46bp,281.0bp) -- (146.46bp,245.0bp) -- (309.58bp,245.0bp) -- cycle;
  \definecolor{strokecol}{rgb}{0.0,0.0,0.0};
  \pgfsetstrokecolor{strokecol}
  \draw (228.02bp,265.8bp) node {input domain};
  \draw (228.02bp,251.8bp) node {e.g. {\tt www.website.com}};
\end{scope}
\begin{scope}
  \definecolor{strokecol}{rgb}{0.0,0.0,0.0};
  \pgfsetstrokecolor{strokecol}
  \draw (112.06bp,195.0bp) -- (-0.02bp,195.0bp) -- (-0.02bp,159.0bp) -- (112.06bp,159.0bp) -- cycle;
  \draw (56.02bp,179.8bp) node {GLRT LSTM};
  \draw (56.02bp,165.8bp) node {(for subdomains)};
\end{scope}
\begin{scope}
  \definecolor{strokecol}{rgb}{0.0,0.0,0.0};
  \pgfsetstrokecolor{strokecol}
  \draw (298.44bp,109.0bp) -- (175.6bp,109.0bp) -- (175.6bp,73.0bp) -- (298.44bp,73.0bp) -- cycle;
  \draw (237.02bp,91.0bp) node {Logistic regression};
\end{scope}
  \end{tikzpicture}
  }
\caption{Overall architecture of our proposed DGA detection model that
incorporates side information.  Input points are split into subdomain, domain,
and TLD.  The subdomain and domain are run through individual RNN+GLRT models,
then the output is combined with the WHOIS data and the one-hot encoded TLD into
the final logistic regression model.}
\label{fig:overall_model}
\end{figure}
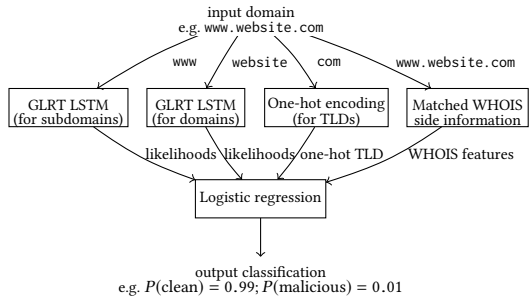
\section{Model Architecture}
\label{sec:model}


Given the effectiveness of deep learning classifiers for character-level DGA
modeling~\cite{woodbridge2016predicting, yu2018character}, we have designed 
our DGA detector on character-level recurrent neural networks
(RNNs)~\cite{karpathy2016visualizing}.  A character-level RNN sequentially
receives characters from a string, updating internal state as each character is
passed in.
Instead of training the RNNs to predict the class of the domain, we instead
train two RNNs to predict the next character in the domain and combine these
predictions via a generalized likelihood ratio test (GLRT).  In addition, our
model also incorporates the WHOIS side information discussed in the previous
section via model stacking.  This allows us to achieve significantly better
performance on more difficult DGA families.

Overall, our model is a logistic regression classifier built on the output of
four different models:

\begin{enumerate}
  \item A character-level RNN GLRT model built only on the {\it subdomains} in
the training set.

  \item A character-level RNN GLRT model built only on the {\it domains} in the
training set.

  \item One-hot encoded top-level domain features (for the most popular 250
TLDs).

  \item Extracted features from the WHOIS information.
\end{enumerate}

The overall architecture of the model can be seen in
Figure~\ref{fig:overall_model}.  In the following subsections we describe the
details of the full model.

\subsection{Character-level RNN GLRT}

The core of the model is the character-level RNN that uses the generalized
likelihood ratio test to classify a domain or a subdomain as DGA or non-DGA.
Previous approaches and other uses of RNNs often predict the class of the output
directly~\cite{woodbridge2016predicting, graves2005framewise};
however, this only allows backpropagation of the error signal at the end of the
entire sequence, which can slow the learning process.

\begin{figure}[b!]
  \centering
  {\small
  \begin{tikzpicture}[scale=0.6]
    \input{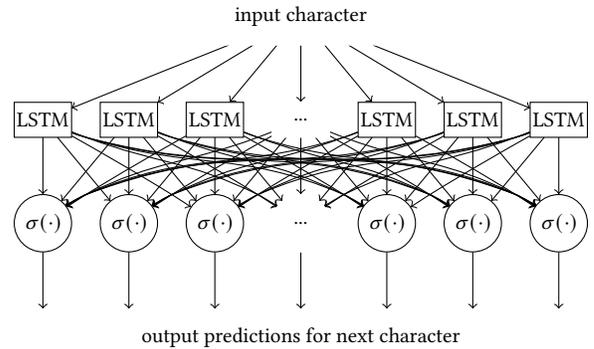}
  \end{tikzpicture}
  }
  \caption{The individual RNN model architecture.  The model takes a one-hot
encoded character as input (along with its current hidden state) in order to
predict the next character in the sequence.}
  \label{fig:lstm}
\end{figure}

Therefore, we build one RNN on each class in the input dataset (in our case,
there are only two classes: {\it DGA} and {\it non-DGA}).  Each input
sequence is converted to a one-hot character encoding, and the label or expected
output of the RNN for each time step is the one-hot encoding of the {\it next}
character in the sequence.  This means the RNN is trained to predict the next
character in the sequence.  Thus, backpropagation can be done at every timestep,
instead of waiting until the end of the sequence to compare the output of the
RNN with the desired label.  Our model's architecture is a single LSTM
layer~\cite{hochreiter1997long} followed by a single dense layer, pictured in
Figure~\ref{fig:lstm}.  We use LSTMs to help avoid the vanishing and exploding
gradient phenomenons~\cite{pascanu2013difficulty}. Although it is possible to
build a more complex network, we found that this provides a good balance between
training time and the accuracy of the model.

In order to perform the one-hot encoding, we first build a dictionary $D$ on the
entire training set, including the `unknown' character {\tt '?'}.  If a character
is encountered at prediction time that is not in $D$, then it is encoded as {\tt
'?'}.

We use the categorical cross-entropy~\cite{goodfellow2016deep} for the loss
function.  Then, during prediction, at each time step $i$ for the input $x$, the
output of the model $\theta$ is a probability $p(x_i | (x_0, \ldots, x_{i - 1}),
\theta)$ and with this we can construct an estimate of the likelihood of the
point $x$ arising from the model $\theta$:

\begin{equation}
p(x | \theta) = \prod_{i} p(x_i | (x_0, \ldots, x_{i - 1}), \theta).
\end{equation}

For the generalized likelihood ratio test~\cite{neyman1933ix}, if we calculated
both likelihood estimates $p(x | \theta_{\textrm{non-dga}})$ and $p(x |
\theta_{\textrm{dga}})$, we could then set a threshold $\eta$ and compute

\begin{equation}
\Lambda(x) = \frac{p(x | \theta_{\textrm{dga}})}{p(x |
\theta_{\textrm{non-dga}})},
\end{equation}

\noindent and if $\eta > \Lambda(x)$, we classify the point as a DGA domain;
otherwise, we classify the point as non-DGA.  The value of $\eta$ can be swept in
order to control the false positive and true positive rate.  $\eta$ is directly
related to the typical posterior probability of a classifier; in fact, if we
normalize the likelihood estimates we can produce a posterior probability of $x$
being a DGA domain:

\begin{equation}
p(\theta_{\textrm{dga}} | x) = \frac{p(x |
\theta_{\textrm{dga}})}{p(x | \theta_{\textrm{non-dga}}) + p(x |
\theta_{\textrm{dga}})}.
\end{equation}

Then, setting a threshold for $p(\theta_{\textrm{dga}})$ is reducible to
setting a GLRT threshold $\eta$.

For our DGA classifier, we build two separate RNN-GLRT models as described
above: one on the {\it subdomains} of our training set, and one on the {\it
domains}.  Each of these two models, in turn, contains a separately-trained LSTM
RNN, whose outputs are combined to perform the GLRT as shown above.

As input to the logistic regression model, we extract six features from each
RNN-GLRT model, giving a total of twelve features.  The features are listed
below.

\begin{itemize}
  \itemsep 5pt
  \item A boolean feature indicating whether a domain or subdomain could be
extracted from the input domain $x$.

  \item The likelihood estimate $p(x | \theta_{\mathrm{non-dga}})$.

  \item The likelihood estimate $p(x | \theta_{\mathrm{dga}})$.

  \item The posterior probability $p(\theta_{\mathrm{non-dga}} | x)$.

  \item The posterior probability $p(\theta_{\mathrm{dga}} | x)$.

  \item The likelihood ratio $\Lambda(x)$.
\end{itemize}

Since we are extracting the likelihood estimates and posterior probabilities
into a logistic regression model, we actually have no need to select a threshold
$\eta$---that is only needed for a standalone GLRT LSTM model.  Instead, in our
combined model, the logistic regression will learn directly from the
probabilities and likelihoods.

\vspace*{-0.4em}
\subsection{Top-level domain features}

Since TLDs are so short (usually two or three characters), it is excessive to
train an RNN on them.  Therefore, we use a one-hot encoding of the TLD, matching
against the 249 most frequent TLDs in our training dataset; if there is no
match, the TLD is encoded as `other', giving a total of 250 binary features out
of the TLD.

In order to perform the conversion, we used the TLD list available from
\url{http://publicsuffix.org}.  The most common TLDs in our dataset were {\tt
.com}, {\tt .org}, {\tt .ru}, {\tt .net}, and {\tt .info}.  We found that the
{\tt .ru}, {\tt .info}, {\tt .biz}, and {\tt .cc} TLDs contained significantly
higher concentrations of DGA domains, with each of those TLDs containing
at least 3 times as many DGA as non-DGA domains.  Since we have split these
into separate features, we can expect our model to learn which TLDs domain
generation algorithms are more likely to use.

\vspace*{-0.4em}
\subsection{WHOIS side information}


The WHOIS data makes up the rest of the input to the logistic regression model.
It is concatenated with the RNN-GLRT features for the domain, the RNN-GLRT
features for the subdomain, and the one-hot encoded TLD features.

Before all of these features are fed into the logistic regression
model, we perform whitening via PCA for decorrelation and
scaling~\cite{kessy2015optimal}.  This step can improve the performance of the
model, although it generally also makes interpretability more difficult.

\vspace*{-0.4em}
\subsection{Computational concerns}

Recurrent neural networks, especially those with complex memory cells like
LSTMs, are well-known to be time-consuming to train~\cite{li2015fpga,
doetsch2014fast}.  Our model is not exempt from this; for large datasets, it may
take many hours to train\footnote{Our training was conducted on a high-end
consumer-grade system with a single GPU.}.  However, in practice this is not a
concern---a single forward pass through the model for classification is
comparatively very fast, and once our model is trained, there are no
computational difficulties with deployment in a low-latency
detection system.  This means that the model can be, e.g., deployed into a
consumer endpoint security product without problems.

\vspace*{-0.6em}
\section{Adversarial samples}

In recent years, the phenomenon of {\it adversarial samples} has surfaced in the
deep learning community~\cite{szegedy2013intriguing, goodfellow2014explaining}.
In essence, a malicious actor could take a sample that was correctly classified
by the model, perturb the input slightly, and the perturbed sample would be
misclassified.  When images are used, these perturbations are often invisible to
the eye.  These adversarial attacks have been successfully applied to fields
outside of images, including audio~\cite{carlini2018audio} and malware
classification~\cite{grosse2017adversarial}.  Though there are some defense
mechanisms that have been developed~\cite{papernot2016distillation,
feinman2017detecting}, many of these are later found to be
circumventable~\cite{carlini2017adversarial}.

Given that adversarial samples are not limited to images, it is reasonable to
believe that neural network-based DGA detectors could also suffer from this
vulnerability.  In our situation a malicious actor would wish to take a domain
that is detected as from a DGA and have it labeled as a non-DGA domain.  It
would be very straightforward to perform an attack like the Fast Gradient Sign
Method~\cite{szegedy2013intriguing} to modify the characters in a domain name.
In fact, it is not (generally) important to DGA authors what the domain name
looks like, so there is no cost to modify the letters of the domain.

Such a technique would likely prove effective against an approach that only
incorporated the domain name itself.  However, note that our model also
incorporates domain registration side information from WHOIS.  Although a
malware author can change the domain name they are using at will and nearly
arbitrarily, it is significantly more difficult to cause the WHOIS registration
information (such as registration date) to have specific values.  To do that, a
malware author might need to register a domain perhaps months in advance and
host a clean website on it, which is both expensive and time-consuming.  Thus,
it would be more difficult for a malware author to work around our proposed
model.

\section{Experiments}

\begin{figure*}[t!]
\begin{subfigure}[b]{0.48\textwidth}
  \centering
  \includegraphics[width=0.95\textwidth]{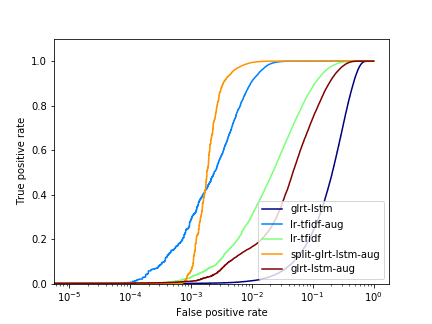}
  \vspace*{-0.8em}
  \caption{ROC curves for {\tt matsnu} family.}
  \label{fig:matsnu_roc}
\end{subfigure}
\begin{subfigure}[b]{0.48\textwidth}
  \centering
  \includegraphics[width=0.95\textwidth]{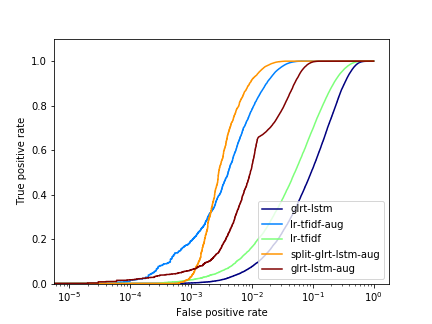}
  \vspace*{-0.8em}
  \caption{ROC curves for {\tt rovnix} family.}
  \label{fig:rovnix_roc}
\end{subfigure}
\begin{subfigure}[b]{0.48\textwidth}
  \vspace*{-0.3em}
  \centering
  \includegraphics[width=0.95\textwidth]{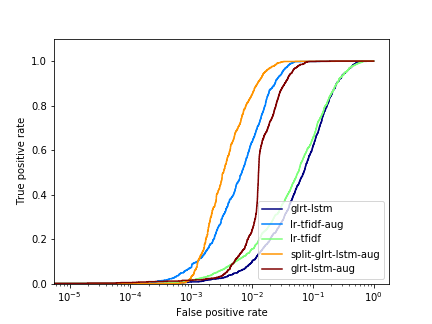}
  \vspace*{-0.8em}
  \caption{ROC curves for {\tt gozi} family.}
  \label{fig:gozi_roc}
\end{subfigure}
\begin{subfigure}[b]{0.48\textwidth}
  \vspace*{-0.3em}
  \centering
  \includegraphics[width=0.95\textwidth]{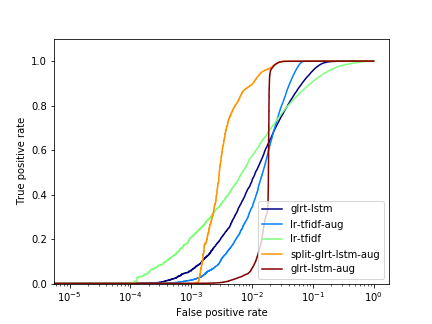}
  \vspace*{-0.8em}
  \caption{ROC curves for {\tt banjori} family.}
  \label{fig:banjori_roc}
\end{subfigure}
\vspace*{-0.8em}
\caption{ROC curves for different families.}
\label{fig:rocs}
\vspace*{-0.8em}
\end{figure*}

\begin{table*}[htb]
\begin{center}
\begin{tabular}{lcccccc}
\toprule
{\bf family} & {\bf $\bar{s}(x)$} & {\bf \ \ \ \ lr-tfidf \ \ \ \ } & {\bf \ \
lr-tfidf-aug \ \ } & {\bf
\ \ \ \ glrt-lstm \ \ \ \ } & {\bf \ \ glrt-lstm-aug \ \ } & {\bf split-glrt-lstm-aug} \\
\midrule
matsnu                      & 332.6 & 0.581 & 0.824 & 0.501 & 0.546 & {\bf 0.891} \\
rovnix         & 284.2 & 0.540 & 0.758 & 0.515 & 0.635 & {\bf 0.805} \\
gozi                        & 222.2 & 0.540 & 0.684 & 0.520 & 0.548 & {\bf 0.773} \\
banjori                     & 173.9 & 0.701 & 0.585 & 0.634 & 0.508 & {\bf 0.808} \\
suppobox  & 152.5 & 0.509 & 0.505 & 0.579 & {\bf 0.798} & 0.568 \\
volatile               & 112.4 & 0.605 & 0.498 & 0.818 & 0.850 & {\bf 0.958} \\
others\_dga\_b    &  90.4 & 0.649 & 0.502 & {\bf 0.704} & 0.604 & 0.677 \\
\midrule
beebone        &  66.2 & 0.498 & 0.498 & 0.498 & 0.498 & {\bf 0.749} \\
ramdo          &  51.4 & 0.902 & 0.498 & 0.973 & 0.498 & {\bf 0.980} \\
qakbot                      &  43.2 & 0.940 & 0.498 & {\bf 0.966} & 0.914 & 0.951 \\
pykspa                      &  40.6 & 0.777 & 0.498 & 0.911 & 0.866 & {\bf 0.957} \\
murofet                     &  38.7 & 0.975 & 0.498 & 0.960 & 0.498 & {\bf 0.994} \\
ranbyus                     &  35.7 & 0.940 & 0.498 & {\bf 0.963} & 0.815 & 0.791 \\
tinba                       &  31.6 & 0.859 & 0.498 & {\bf 0.968} & 0.655 & 0.867 \\
necurs                      &  31.3 & 0.818 & 0.498 & {\bf 0.813} & 0.514 & 0.581 \\
new\_goz                    &  23.3 & 0.863 & 0.498 & {\bf 0.993} & 0.498 & 0.990 \\
\midrule
shiotob                     &  21.0 & 0.653 & 0.498 & 0.894 & 0.846 & {\bf 0.902} \\
locky                       &  20.8 & {\bf 0.849} & 0.498 & 0.751 & 0.498 & 0.557 \\
proslikefan                &  20.8 & 0.647 & 0.498 & {\bf 0.802} & 0.600 & 0.665 \\
conficker      &  19.9 & 0.609 & 0.498 & {\bf 0.836} & 0.619 & 0.649 \\
chinad                      &  19.6 & 0.786 & 0.498 & 0.952 & 0.498 & {\bf 0.979} \\
dyre   &   6.5 & 0.548 & 0.498 & 0.779 & 0.498 & {\bf 0.974} \\
\bottomrule
\end{tabular}
\end{center}
\caption{Partial AUC (FPR <= 0.01) performance numbers for selected different
DGA families, with a focus on those with either high or low average smashword
score.  Our {\bf split-glrt-lstm-aug} model performs best on those DGA families
with high average smashword score.}
\label{tab:aucs}
\vspace*{-2.0em}
\end{table*}





The most important situation for any DGA detection model is when it encounters
an entirely new DGA family that it has never seen before.  This is the situation
that we focus on in our experiments, since it reflects the real-world `zero-day'
situation.  We compare our model to several baselines that reflect the
state-of-the-art for machine learning systems that do not use network traffic
data.

{\bf Leave-one-out models.}  To simulate the situation where a DGA family has
not been seen, we validate the performance of our DGA detection model by
performing {\it leave-one-out} experiments, where we train the model on all DGA
families except one, and then the test set consists entirely of the left-out DGA
family combined with some never-before-seen non-DGA domains.  This shows us how
well the model is able to generalize to unseen DGA family types.

{\bf Dataset details.}  Our collected dataset, as described in
Section~\ref{sec:families}, includes 41 DGA families plus non-DGA data, totaling
2.3 million domain names (1.01 million non-DGA, 1.28 million DGA).
Of these 41 DGA families, many with high average smashword score have been
specifically mentioned in related work as difficult.  The LSTM model of
Woodbridge et~al.~\cite{woodbridge2016predicting} is specifically shown to
perform very poorly on the {\tt matsnu}, {\tt suppobox}, and {\tt
beebone} families, each of which have above average to very large average
smashword scores.  Mac et~al.~\cite{mac2017dga} claim that {\tt matsnu} is not
differentiable from non-DGA domains at all, and show very poor performance on all
their surveyed algorithms for the {\tt nymaim} DGA, which is very similar to the
{\tt gozi} DGA that we use here.  Because our model has been specifically
designed to focus on DGA families that are understood to be more difficult, we
will focus on these families.

{\bf Baseline models.}  We wish to compare the performance of our proposed model
with existing and baseline approaches.  Therefore, we compare our model with
four other models, which we now introduce.  Two of these are simple baseline
models, with and without WHOIS information, and the other two are based on LSTM
architectures that represent the most closely related state-of-the-art work of
Woodbridge et~al.~\cite{woodbridge2016predicting}.

\begin{itemize}
  \itemsep 5pt
  \item {\bf lr-tfidf}: logistic regression on TF-IDF features extracted
  from the domain name using 2-grams.\footnote{We did not use 3-grams, because
  the memory usage on our system was too large.}  Any WHOIS side information is
  not used here, so this model presents a reasonable baseline using only the
  domain name.

  \item {\bf lr-tfidf-aug}: logistic regression on TF-IDF features extracted
  from the domain name using 2-grams, and augmented with the WHOIS features.
  This is a reasonable baseline for classification using both the domain name
  and the side information (WHOIS features).

  \item {\bf glrt-lstm}: a GLRT LSTM model built only on the full domain name
  (no side information).  This can be considered to be a slight improvement over
  the model of Woodbridge et~al. \cite{woodbridge2016predicting} due to the use
  of the GLRT.

  \item {\bf glrt-lstm-aug}: a GLRT LSTM model built only on the full domain
  name, and then used as input to a logistic regression model, with the WHOIS
  features augmented.

\end{itemize}

{\bf Our model.}  We refer to our model as the {\bf split-glrt-lstm-aug} model;
this is the model from Section \ref{sec:model}.

{\bf Training and implementation details.}  The {\bf lr-tfidf} and {\bf
lr-tfidf-aug} models were implemented with {\tt
scikit-learn}~\cite{pedregosa2011scikit}, and the three LSTM-based models were
implemented with Keras~\cite{chollet2015keras} using the TensorFlow
backend~\cite{abadi2016tensorflow}.  Each LSTM model used 500 LSTM units and was
trained for 100 epochs (passes over the dataset) with early stopping using the
RMSprop optimizer, with dropout of 0.2.  With our setup (one nVidia GeForce GTX
TITAN X), each LSTM model took approximately 8-10 hours to train.  In our
experiments, we found that changing the optimizer made little difference to the
resulting model, and we found that increasing or decreasing the number of LSTM
units decreased performance slightly.  Overall, our model seemed to be
relatively robust to hyperparameter choice.

Figure~\ref{fig:rocs} shows receiver operating characteristic curves (ROC
curves) on the four datasets with highest smashword score.  We can see in the
figures that the {\bf split-glrt-lstm-aug} model (our proposed model)
outperforms each of the other models, providing better performance at lower
false positive rates.  For instance, on the difficult {\tt matsnu} family, when
the false positive rate is chosen to be 0.5\%, the {\bf split-glrt-lstm-aug}
model operates at a true positive rate of
95\%,
whereas the next best model ({\bf lr-tfidf-aug}) operates at a true positive
rate of only
70\%.

In typical application scenarios, we typically care only about running our
classifier at false positive rates less than or equal to 1\% (FPR $\le 0.01$).
Therefore, we study the performance of the classifiers using the {\it partial
AUC} \cite{mcclish1989analyzing} measure, which is the standard
area-under-the-curve (AUC) measure specific to false positive rates less than a
given threshold.  In Table \ref{tab:aucs}, we show the partial AUC of each model
for each leave-one-out family experiment, sorted by decreasing $\bar{s}(\cdot)$.

On the most difficult families (with large $\bar{s}(\cdot)$), the proposed {\bf
split-glrt-lstm-aug} reliably and significantly outperforms all other compared
models.  This is the region of most interest in our work, as these families are
difficult to detect---even with WHOIS data.  Each of these difficult families
generates domains that resemble English words; see Table~\ref{tab:families}.
Note that the {\bf lr-tfidf-aug} model and {\bf glrt-lstm-aug} models both have
access to the WHOIS features; however, only {\bf split-glrt-lstm-aug} is able to
take advantage of these to provide good performance for families with high
$\bar{s}(\cdot)$.

For `easier' families with lower $\bar{s}(\cdot)$, where the generated domains
typically look more like random characters, classification can be performed more
reliably with only the text of the domain itself; thus, the {\bf glrt-lstm}
model is dominant in this regime.
%
%

Overall, we see that our model is successful in detecting DGA-generated domains
that resemble English words.  The model appears to generalize well to different
families, given the nature of our leave-one-out experiments.

\section{Conclusion}


In this paper we have considered the problem of DGA domain detection.  We
introduced a measure of complexity for DGA families called the {\it smashword
score}, which reflects how closely a DGA's generated domains resemble English
words.  Because DGA families with higher smashword scores have typically posed
greater difficulty for detection, we build a novel machine learning model
consisting of recurrent neural networks (RNNs) using the generalized likelihood
ratio test (GLRT), and augment these models with a logistic regression model
that also includes side information such as WHOIS information.

This combined model notably outperforms existing state-of-the-art approaches on
DGA families with high smashword score, such as the difficult {\tt matsnu} and
{\tt suppobox} families. We believe that this model could be used as either a
standalone model or as a part of a larger DGA detection system that could also
incorporate network traffic, such as something more like the Pleiades system
\cite{antonakakis2012throw}.

There is room for future improvement in our work. The model we have used is
specialized for DGA families based on English words, and therefore can be less
effective for those DGA families that do not look like natural domain names.
Thus, in a production environment or in an improved system, our model could be
ensembled with other techniques that are more effective for DGA families with
lower smashword scores.

In a future work we would also like to explore multilingual approaches to tackle 
new families that may use non-English dictionaries and expand our side information features.


%
\bibliographystyle{ACM-Reference-Format}
\bibliography{paper}


\begin{thebibliography}{47}


\ifx \showCODEN    \undefined \def \showCODEN     #1{\unskip}     \fi
\ifx \showDOI      \undefined \def \showDOI       #1{#1}\fi
\ifx \showISBNx    \undefined \def \showISBNx     #1{\unskip}     \fi
\ifx \showISBNxiii \undefined \def \showISBNxiii  #1{\unskip}     \fi
\ifx \showISSN     \undefined \def \showISSN      #1{\unskip}     \fi
\ifx \showLCCN     \undefined \def \showLCCN      #1{\unskip}     \fi
\ifx \shownote     \undefined \def \shownote      #1{#1}          \fi
\ifx \showarticletitle \undefined \def \showarticletitle #1{#1}   \fi
\ifx \showURL      \undefined \def \showURL       {\relax}        \fi
\providecommand\bibfield[2]{#2}
\providecommand\bibinfo[2]{#2}
\providecommand\natexlab[1]{#1}
\providecommand\showeprint[2][]{arXiv:#2}

\bibitem[\protect\citeauthoryear{Abadi, Barham, Chen, Chen, Davis, Dean, Devin,
  Ghemawat, Irving, Isard, et~al\mbox{.}}{Abadi et~al\mbox{.}}{2016}]%
        {abadi2016tensorflow}
\bibfield{author}{\bibinfo{person}{Mart{\'\i}n Abadi}, \bibinfo{person}{Paul
  Barham}, \bibinfo{person}{Jianmin Chen}, \bibinfo{person}{Zhifeng Chen},
  \bibinfo{person}{Andy Davis}, \bibinfo{person}{Jeffrey Dean},
  \bibinfo{person}{Matthieu Devin}, \bibinfo{person}{Sanjay Ghemawat},
  \bibinfo{person}{Geoffrey Irving}, \bibinfo{person}{Michael Isard},
  {et~al\mbox{.}}} \bibinfo{year}{2016}\natexlab{}.
\newblock \showarticletitle{{TensorFlow}: a system for large-scale machine
  learning}. In \bibinfo{booktitle}{\emph{Proceedings of the 12th USENIX
  conference on Operating Systems Design and Implementation}}. USENIX
  Association, \bibinfo{pages}{265--283}.
\newblock


\bibitem[\protect\citeauthoryear{Anderson, Woodbridge, and Filar}{Anderson
  et~al\mbox{.}}{2016}]%
        {deepdga16}
\bibfield{author}{\bibinfo{person}{Hyrum~S. Anderson},
  \bibinfo{person}{Jonathan Woodbridge}, {and} \bibinfo{person}{Bobby Filar}.}
  \bibinfo{year}{2016}\natexlab{}.
\newblock \showarticletitle{{DeepDGA: Adversarially-Tuned Domain Generation and
  Detection}}. In \bibinfo{booktitle}{\emph{Proceedings of the 2016 ACM
  Workshop on Artificial Intelligence and Security}}
  \emph{(\bibinfo{series}{AISec '16})}. \bibinfo{publisher}{ACM},
  \bibinfo{address}{New York, NY, USA}, \bibinfo{pages}{13--21}.
\newblock
\showISBNx{978-1-4503-4573-6}
\urldef\tempurl%
\url{https://doi.org/10.1145/2996758.2996767}
\showDOI{\tempurl}


\bibitem[\protect\citeauthoryear{Antonakakis, Perdisci, Nadji, Vasiloglou,
  Abu-Nimeh, Lee, and Dagon}{Antonakakis et~al\mbox{.}}{2012}]%
        {antonakakis2012throw}
\bibfield{author}{\bibinfo{person}{Manos Antonakakis}, \bibinfo{person}{Roberto
  Perdisci}, \bibinfo{person}{Yacin Nadji}, \bibinfo{person}{Nikolaos
  Vasiloglou}, \bibinfo{person}{Saeed Abu-Nimeh}, \bibinfo{person}{Wenke Lee},
  {and} \bibinfo{person}{David Dagon}.} \bibinfo{year}{2012}\natexlab{}.
\newblock \showarticletitle{{From throw-away traffic to bots: detecting the
  rise of DGA-based malware}}. In \bibinfo{booktitle}{\emph{Proceedings of the
  21st USENIX conference on Security symposium}}. USENIX Association,
  \bibinfo{pages}{24--24}.
\newblock


\bibitem[\protect\citeauthoryear{Aviv and Haeberlen}{Aviv and
  Haeberlen}{2011}]%
        {avivbotnet11}
\bibfield{author}{\bibinfo{person}{Adam~J. Aviv} {and} \bibinfo{person}{Andreas
  Haeberlen}.} \bibinfo{year}{2011}\natexlab{}.
\newblock \showarticletitle{Challenges in Experimenting with Botnet Detection
  Systems}. In \bibinfo{booktitle}{\emph{Proceedings of the 4th Conference on
  Cyber Security Experimentation and Test}} \emph{(\bibinfo{series}{CSET'11})}.
  \bibinfo{publisher}{USENIX Association}, \bibinfo{address}{Berkeley, CA,
  USA}, \bibinfo{pages}{6}.
\newblock
\urldef\tempurl%
\url{http://dl.acm.org/citation.cfm?id=2027999.2028005}
\showURL{%
\tempurl}


\bibitem[\protect\citeauthoryear{Bilge, Kirda, Kruegel, and Balduzzi}{Bilge
  et~al\mbox{.}}{2011}]%
        {bilge2011exposure}
\bibfield{author}{\bibinfo{person}{Leyla Bilge}, \bibinfo{person}{Engin Kirda},
  \bibinfo{person}{Christopher Kruegel}, {and} \bibinfo{person}{Marco
  Balduzzi}.} \bibinfo{year}{2011}\natexlab{}.
\newblock \showarticletitle{{EXPOSURE: Finding Malicious Domains Using Passive
  DNS Analysis}}. In \bibinfo{booktitle}{\emph{Proceedings of the 18th Annual
  Network and Distributed System Security Syposium (NDSS 2011)}}.
\newblock


\bibitem[\protect\citeauthoryear{Canali, Cova, Vigna, and Kruegel}{Canali
  et~al\mbox{.}}{2011}]%
        {canali2011prophiler}
\bibfield{author}{\bibinfo{person}{Davide Canali}, \bibinfo{person}{Marco
  Cova}, \bibinfo{person}{Giovanni Vigna}, {and} \bibinfo{person}{Christopher
  Kruegel}.} \bibinfo{year}{2011}\natexlab{}.
\newblock \showarticletitle{Prophiler: a fast filter for the large-scale
  detection of malicious web pages}. In \bibinfo{booktitle}{\emph{Proceedings
  of the 20th International World Wide Web Conference (WWW 2011)}}. ACM,
  \bibinfo{pages}{197--206}.
\newblock


\bibitem[\protect\citeauthoryear{Carlini and Wagner}{Carlini and
  Wagner}{2017}]%
        {carlini2017adversarial}
\bibfield{author}{\bibinfo{person}{Nicholas Carlini} {and}
  \bibinfo{person}{David Wagner}.} \bibinfo{year}{2017}\natexlab{}.
\newblock \showarticletitle{Adversarial examples are not easily detected:
  Bypassing ten detection methods}. In \bibinfo{booktitle}{\emph{Proceedings of
  the 10th ACM Workshop on Artificial Intelligence and Security}}. ACM,
  \bibinfo{pages}{3--14}.
\newblock


\bibitem[\protect\citeauthoryear{Carlini and Wagner}{Carlini and
  Wagner}{2018}]%
        {carlini2018audio}
\bibfield{author}{\bibinfo{person}{Nicholas Carlini} {and}
  \bibinfo{person}{David Wagner}.} \bibinfo{year}{2018}\natexlab{}.
\newblock \showarticletitle{Audio adversarial examples: Targeted attacks on
  speech-to-text}.
\newblock \bibinfo{journal}{\emph{arXiv preprint arXiv:1801.01944}}
  (\bibinfo{year}{2018}).
\newblock


\bibitem[\protect\citeauthoryear{Chollet et~al\mbox{.}}{Chollet
  et~al\mbox{.}}{2015}]%
        {chollet2015keras}
\bibfield{author}{\bibinfo{person}{Fran\c{c}ois Chollet} {et~al\mbox{.}}}
  \bibinfo{year}{2015}\natexlab{}.
\newblock \bibinfo{title}{Keras}.
\newblock \bibinfo{howpublished}{\url{https://keras.io}}.
\newblock


\bibitem[\protect\citeauthoryear{{Council of European Union}}{{Council of
  European Union}}{2016}]%
        {gdpr}
\bibfield{author}{\bibinfo{person}{{Council of European Union}}.}
  \bibinfo{year}{2016}\natexlab{}.
\newblock \bibinfo{title}{{Council regulation ({EU}) no. 2016/679 (General Data
  Protection Regulation)}}.
\newblock
\newblock
\newblock
\shownote{\url{https://eur-lex.europa.eu/legal-content/EN/TXT/?uri=uriserv:OJ.L_.2016.119.01.0001.01.ENG}.}


\bibitem[\protect\citeauthoryear{Daigle}{Daigle}{2004}]%
        {rfc3912}
\bibfield{author}{\bibinfo{person}{L. Daigle}.}
  \bibinfo{year}{2004}\natexlab{}.
\newblock \bibinfo{booktitle}{\emph{{WHOIS Protocol Specification}}}.
\newblock \bibinfo{type}{{RFC}} 3912. \bibinfo{institution}{{RFC Editor}}.
  \bibinfo{pages}{1--4} pages.
\newblock
\urldef\tempurl%
\url{https://www.rfc-editor.org/rfc/rfc3912.txt}
\showURL{%
\tempurl}


\bibitem[\protect\citeauthoryear{Doetsch, Kozielski, and Ney}{Doetsch
  et~al\mbox{.}}{2014}]%
        {doetsch2014fast}
\bibfield{author}{\bibinfo{person}{Patrick Doetsch}, \bibinfo{person}{Michal
  Kozielski}, {and} \bibinfo{person}{Hermann Ney}.}
  \bibinfo{year}{2014}\natexlab{}.
\newblock \showarticletitle{Fast and robust training of recurrent neural
  networks for offline handwriting recognition}. In
  \bibinfo{booktitle}{\emph{Proceedings of the 2014 14th International
  Conference on Frontiers in Handwriting Recognition (ICFHR)}}. IEEE,
  \bibinfo{pages}{279--284}.
\newblock


\bibitem[\protect\citeauthoryear{Feinman, Curtin, Shintre, and Gardner}{Feinman
  et~al\mbox{.}}{2017}]%
        {feinman2017detecting}
\bibfield{author}{\bibinfo{person}{Reuben Feinman}, \bibinfo{person}{Ryan~R
  Curtin}, \bibinfo{person}{Saurabh Shintre}, {and} \bibinfo{person}{Andrew~B
  Gardner}.} \bibinfo{year}{2017}\natexlab{}.
\newblock \showarticletitle{Detecting adversarial samples from artifacts}.
\newblock \bibinfo{journal}{\emph{arXiv preprint arXiv:1703.00410}}
  (\bibinfo{year}{2017}).
\newblock


\bibitem[\protect\citeauthoryear{Goodfellow, Bengio, Courville, and
  Bengio}{Goodfellow et~al\mbox{.}}{2016}]%
        {goodfellow2016deep}
\bibfield{author}{\bibinfo{person}{Ian Goodfellow}, \bibinfo{person}{Yoshua
  Bengio}, \bibinfo{person}{Aaron Courville}, {and} \bibinfo{person}{Yoshua
  Bengio}.} \bibinfo{year}{2016}\natexlab{}.
\newblock \bibinfo{booktitle}{\emph{{Deep Learning}}}.
\newblock \bibinfo{publisher}{MIT Press Cambridge}.
\newblock


\bibitem[\protect\citeauthoryear{Goodfellow, Shlens, and Szegedy}{Goodfellow
  et~al\mbox{.}}{2014}]%
        {goodfellow2014explaining}
\bibfield{author}{\bibinfo{person}{Ian~J Goodfellow}, \bibinfo{person}{Jonathon
  Shlens}, {and} \bibinfo{person}{Christian Szegedy}.}
  \bibinfo{year}{2014}\natexlab{}.
\newblock \showarticletitle{Explaining and Harnessing Adversarial Examples}.
\newblock \bibinfo{journal}{\emph{arXiv preprint arXiv:1412.6572}}
  (\bibinfo{year}{2014}).
\newblock


\bibitem[\protect\citeauthoryear{Graves and Schmidhuber}{Graves and
  Schmidhuber}{2005}]%
        {graves2005framewise}
\bibfield{author}{\bibinfo{person}{Alex Graves} {and}
  \bibinfo{person}{J{\"u}rgen Schmidhuber}.} \bibinfo{year}{2005}\natexlab{}.
\newblock \showarticletitle{{Framewise phoneme classification with
  bidirectional LSTM and other neural network architectures}}.
\newblock \bibinfo{journal}{\emph{Neural Networks}} \bibinfo{volume}{18},
  \bibinfo{number}{5-6} (\bibinfo{year}{2005}), \bibinfo{pages}{602--610}.
\newblock


\bibitem[\protect\citeauthoryear{Grosse, Papernot, Manoharan, Backes, and
  McDaniel}{Grosse et~al\mbox{.}}{2017}]%
        {grosse2017adversarial}
\bibfield{author}{\bibinfo{person}{Kathrin Grosse}, \bibinfo{person}{Nicolas
  Papernot}, \bibinfo{person}{Praveen Manoharan}, \bibinfo{person}{Michael
  Backes}, {and} \bibinfo{person}{Patrick McDaniel}.}
  \bibinfo{year}{2017}\natexlab{}.
\newblock \showarticletitle{Adversarial examples for malware detection}. In
  \bibinfo{booktitle}{\emph{European Symposium on Research in Computer
  Security}}. Springer, \bibinfo{pages}{62--79}.
\newblock


\bibitem[\protect\citeauthoryear{Hochreiter and Schmidhuber}{Hochreiter and
  Schmidhuber}{1997}]%
        {hochreiter1997long}
\bibfield{author}{\bibinfo{person}{Sepp Hochreiter} {and}
  \bibinfo{person}{J{\"u}rgen Schmidhuber}.} \bibinfo{year}{1997}\natexlab{}.
\newblock \showarticletitle{Long short-term memory}.
\newblock \bibinfo{journal}{\emph{Neural computation}} \bibinfo{volume}{9},
  \bibinfo{number}{8} (\bibinfo{year}{1997}), \bibinfo{pages}{1735--1780}.
\newblock


\bibitem[\protect\citeauthoryear{{ICANN}}{{ICANN}}{2018}]%
        {icann-gdpr}
\bibfield{author}{\bibinfo{person}{{ICANN}}.} \bibinfo{year}{2018}\natexlab{}.
\newblock \bibinfo{title}{{Data Protection/Privacy Issues}}.
\newblock
\newblock
\newblock
\shownote{\url{https://www.icann.org/dataprotectionprivacy}.}


\bibitem[\protect\citeauthoryear{Jiang, Cao, Jin, Li, and Zhang}{Jiang
  et~al\mbox{.}}{2010}]%
        {dnsgraph10}
\bibfield{author}{\bibinfo{person}{Nan Jiang}, \bibinfo{person}{Jin Cao},
  \bibinfo{person}{Yu Jin}, \bibinfo{person}{Li~Erran Li}, {and}
  \bibinfo{person}{Zhi-Li Zhang}.} \bibinfo{year}{2010}\natexlab{}.
\newblock \showarticletitle{{Identifying Suspicious Activities Through DNS
  Failure Graph Analysis}}. In \bibinfo{booktitle}{\emph{Proceedings of the The
  18th IEEE International Conference on Network Protocols}}
  \emph{(\bibinfo{series}{ICNP '10})}. \bibinfo{publisher}{IEEE Computer
  Society}, \bibinfo{address}{Washington, DC, USA}, \bibinfo{pages}{144--153}.
\newblock
\showISBNx{978-1-4244-8644-1}
\urldef\tempurl%
\url{https://doi.org/10.1109/ICNP.2010.5762763}
\showDOI{\tempurl}


\bibitem[\protect\citeauthoryear{Karpathy, Johnson, and Fei-Fei}{Karpathy
  et~al\mbox{.}}{2016}]%
        {karpathy2016visualizing}
\bibfield{author}{\bibinfo{person}{Andrej Karpathy}, \bibinfo{person}{Justin
  Johnson}, {and} \bibinfo{person}{Li Fei-Fei}.}
  \bibinfo{year}{2016}\natexlab{}.
\newblock \showarticletitle{Visualizing and understanding recurrent networks}.
\newblock \bibinfo{journal}{\emph{arXiv preprint arXiv:1506.02078}}
  (\bibinfo{year}{2016}).
\newblock


\bibitem[\protect\citeauthoryear{Kessy, Lewin, and Strimmer}{Kessy
  et~al\mbox{.}}{2018}]%
        {kessy2015optimal}
\bibfield{author}{\bibinfo{person}{Agnan Kessy}, \bibinfo{person}{Alex Lewin},
  {and} \bibinfo{person}{Korbinian Strimmer}.} \bibinfo{year}{2018}\natexlab{}.
\newblock \showarticletitle{Optimal whitening and decorrelation}.
\newblock \bibinfo{journal}{\emph{The American Statistician}}
  (\bibinfo{year}{2018}), \bibinfo{pages}{1--6}.
\newblock


\bibitem[\protect\citeauthoryear{Kountouras, Kintis, Lever, Chen, Nadji, Dagon,
  Antonakakis, and Joffe}{Kountouras et~al\mbox{.}}{2016}]%
        {kountouras2016enabling}
\bibfield{author}{\bibinfo{person}{Athanasios Kountouras},
  \bibinfo{person}{Panagiotis Kintis}, \bibinfo{person}{Chaz Lever},
  \bibinfo{person}{Yizheng Chen}, \bibinfo{person}{Yacin Nadji},
  \bibinfo{person}{David Dagon}, \bibinfo{person}{Manos Antonakakis}, {and}
  \bibinfo{person}{Rodney Joffe}.} \bibinfo{year}{2016}\natexlab{}.
\newblock \showarticletitle{{Enabling network security through active DNS
  datasets}}. In \bibinfo{booktitle}{\emph{International Symposium on Research
  in Attacks, Intrusions, and Defenses}}. Springer, \bibinfo{pages}{188--208}.
\newblock


\bibitem[\protect\citeauthoryear{Leder and Martini}{Leder and Martini}{2009}]%
        {kraken09}
\bibfield{author}{\bibinfo{person}{Felix~S. Leder} {and} \bibinfo{person}{Peter
  Martini}.} \bibinfo{year}{2009}\natexlab{}.
\newblock \showarticletitle{NGBPA Next Generation BotNet Protocol Analysis}. In
  \bibinfo{booktitle}{\emph{Emerging Challenges for Security, Privacy and
  Trust}}, \bibfield{editor}{\bibinfo{person}{Dimitris Gritzalis} {and}
  \bibinfo{person}{Javier Lopez}} (Eds.). \bibinfo{publisher}{Springer Berlin
  Heidelberg}, \bibinfo{address}{Berlin, Heidelberg},
  \bibinfo{pages}{307--317}.
\newblock
\showISBNx{978-3-642-01244-0}


\bibitem[\protect\citeauthoryear{Lever, Walls, Nadji, Dagon, McDaniel, and
  Antonakakis}{Lever et~al\mbox{.}}{2016}]%
        {lever2016domain}
\bibfield{author}{\bibinfo{person}{Chaz Lever}, \bibinfo{person}{Robert Walls},
  \bibinfo{person}{Yacin Nadji}, \bibinfo{person}{David Dagon},
  \bibinfo{person}{Patrick McDaniel}, {and} \bibinfo{person}{Manos
  Antonakakis}.} \bibinfo{year}{2016}\natexlab{}.
\newblock \showarticletitle{{Domain-Z: 28 registrations later measuring the
  exploitation of residual trust in domains}}. In
  \bibinfo{booktitle}{\emph{2016 IEEE Symposium on Security and Privacy
  (S\&P)}}. IEEE, \bibinfo{pages}{691--706}.
\newblock


\bibitem[\protect\citeauthoryear{Li, Wu, Li, Li, Wang, and Qiu}{Li
  et~al\mbox{.}}{2015}]%
        {li2015fpga}
\bibfield{author}{\bibinfo{person}{Sicheng Li}, \bibinfo{person}{Chunpeng Wu},
  \bibinfo{person}{Hai Li}, \bibinfo{person}{Boxun Li}, \bibinfo{person}{Yu
  Wang}, {and} \bibinfo{person}{Qinru Qiu}.} \bibinfo{year}{2015}\natexlab{}.
\newblock \showarticletitle{Fpga acceleration of recurrent neural network based
  language model}. In \bibinfo{booktitle}{\emph{2015 IEEE 23rd Annual
  International Symposium on Field-Programmable Custom Computing Machines
  (FCCM)}}. IEEE, \bibinfo{pages}{111--118}.
\newblock


\bibitem[\protect\citeauthoryear{Luo, Wang, Xu, Yang, Sun, and Wang}{Luo
  et~al\mbox{.}}{2017}]%
        {dgasensor17}
\bibfield{author}{\bibinfo{person}{Xi Luo}, \bibinfo{person}{Liming Wang},
  \bibinfo{person}{Zhen Xu}, \bibinfo{person}{Jing Yang}, \bibinfo{person}{Mo
  Sun}, {and} \bibinfo{person}{Jing Wang}.} \bibinfo{year}{2017}\natexlab{}.
\newblock \showarticletitle{{DGASensor: Fast Detection for DGA-Based
  Malwares}}. In \bibinfo{booktitle}{\emph{Proceedings of the 5th International
  Conference on Communications and Broadband Networking}}
  \emph{(\bibinfo{series}{ICCBN '17})}. \bibinfo{publisher}{ACM},
  \bibinfo{address}{New York, NY, USA}, \bibinfo{pages}{47--53}.
\newblock
\showISBNx{978-1-4503-4861-4}
\urldef\tempurl%
\url{https://doi.org/10.1145/3057109.3057112}
\showDOI{\tempurl}


\bibitem[\protect\citeauthoryear{Ma, Saul, Savage, and Voelker}{Ma
  et~al\mbox{.}}{2009}]%
        {ma2009beyond}
\bibfield{author}{\bibinfo{person}{Justin Ma}, \bibinfo{person}{Lawrence~K.
  Saul}, \bibinfo{person}{Stefan Savage}, {and} \bibinfo{person}{Geoffrey~M.
  Voelker}.} \bibinfo{year}{2009}\natexlab{}.
\newblock \showarticletitle{{Beyond blacklists: learning to detect malicious
  web sites from suspicious URLs}}. In \bibinfo{booktitle}{\emph{Proceedings of
  the 15th ACM SIGKDD International Conference on Knowledge Discovery and Data
  Mining (KDD 2009)}}. ACM, \bibinfo{pages}{1245--1254}.
\newblock


\bibitem[\protect\citeauthoryear{Mac, Tran, Tong, Nguyen, and Tran}{Mac
  et~al\mbox{.}}{2017}]%
        {mac2017dga}
\bibfield{author}{\bibinfo{person}{Hieu Mac}, \bibinfo{person}{Duc Tran},
  \bibinfo{person}{Van Tong}, \bibinfo{person}{Linh~Giang Nguyen}, {and}
  \bibinfo{person}{Hai~Anh Tran}.} \bibinfo{year}{2017}\natexlab{}.
\newblock \showarticletitle{{DGA Botnet Detection Using Supervised Learning
  Methods}}. In \bibinfo{booktitle}{\emph{Proceedings of the Eighth
  International Symposium on Information and Communication Technology}}
  \emph{(\bibinfo{series}{SoICT 2017})}. \bibinfo{publisher}{ACM},
  \bibinfo{address}{New York, NY, USA}, \bibinfo{pages}{211--218}.
\newblock
\showISBNx{978-1-4503-5328-1}
\urldef\tempurl%
\url{https://doi.org/10.1145/3155133.3155166}
\showDOI{\tempurl}


\bibitem[\protect\citeauthoryear{McClish}{McClish}{1989}]%
        {mcclish1989analyzing}
\bibfield{author}{\bibinfo{person}{Donna~K. McClish}.}
  \bibinfo{year}{1989}\natexlab{}.
\newblock \showarticletitle{{Analyzing a portion of the ROC curve}}.
\newblock \bibinfo{journal}{\emph{Medical Decision Making}}
  \bibinfo{volume}{9}, \bibinfo{number}{3} (\bibinfo{year}{1989}),
  \bibinfo{pages}{190--195}.
\newblock


\bibitem[\protect\citeauthoryear{Neyman and Pearson}{Neyman and
  Pearson}{1933}]%
        {neyman1933ix}
\bibfield{author}{\bibinfo{person}{Jerzy Neyman} {and} \bibinfo{person}{Egon~S
  Pearson}.} \bibinfo{year}{1933}\natexlab{}.
\newblock \showarticletitle{{IX. On the problem of the most efficient tests of
  statistical hypotheses}}.
\newblock \bibinfo{journal}{\emph{Phil. Trans. R. Soc. Lond. A}}
  \bibinfo{volume}{231}, \bibinfo{number}{694-706} (\bibinfo{year}{1933}),
  \bibinfo{pages}{289--337}.
\newblock


\bibitem[\protect\citeauthoryear{Papernot, McDaniel, Wu, Jha, and
  Swami}{Papernot et~al\mbox{.}}{2016}]%
        {papernot2016distillation}
\bibfield{author}{\bibinfo{person}{Nicolas Papernot}, \bibinfo{person}{Patrick
  McDaniel}, \bibinfo{person}{Xi Wu}, \bibinfo{person}{Somesh Jha}, {and}
  \bibinfo{person}{Ananthram Swami}.} \bibinfo{year}{2016}\natexlab{}.
\newblock \showarticletitle{Distillation as a defense to adversarial
  perturbations against deep neural networks}. In
  \bibinfo{booktitle}{\emph{Security and Privacy (SP), 2016 IEEE Symposium
  on}}. IEEE, \bibinfo{pages}{582--597}.
\newblock


\bibitem[\protect\citeauthoryear{Pascanu, Mikolov, and Bengio}{Pascanu
  et~al\mbox{.}}{2013}]%
        {pascanu2013difficulty}
\bibfield{author}{\bibinfo{person}{Razvan Pascanu}, \bibinfo{person}{Tomas
  Mikolov}, {and} \bibinfo{person}{Yoshua Bengio}.}
  \bibinfo{year}{2013}\natexlab{}.
\newblock \showarticletitle{On the difficulty of training recurrent neural
  networks}. In \bibinfo{booktitle}{\emph{Proceedings of the 30th International
  Conference on Machine Learning (ICML '13)}}, Vol.~\bibinfo{volume}{28}.
  \bibinfo{publisher}{PMLR}, \bibinfo{address}{Atlanta, Georgia, USA},
  \bibinfo{pages}{1310--1318}.
\newblock


\bibitem[\protect\citeauthoryear{Pedregosa, Varoquaux, Gramfort, Michel,
  Thirion, Grisel, Blondel, Prettenhofer, Weiss, Dubourg,
  et~al\mbox{.}}{Pedregosa et~al\mbox{.}}{2011}]%
        {pedregosa2011scikit}
\bibfield{author}{\bibinfo{person}{Fabian Pedregosa}, \bibinfo{person}{Ga{\"e}l
  Varoquaux}, \bibinfo{person}{Alexandre Gramfort}, \bibinfo{person}{Vincent
  Michel}, \bibinfo{person}{Bertrand Thirion}, \bibinfo{person}{Olivier
  Grisel}, \bibinfo{person}{Mathieu Blondel}, \bibinfo{person}{Peter
  Prettenhofer}, \bibinfo{person}{Ron Weiss}, \bibinfo{person}{Vincent
  Dubourg}, {et~al\mbox{.}}} \bibinfo{year}{2011}\natexlab{}.
\newblock \showarticletitle{{Scikit-learn: Machine learning in Python}}.
\newblock \bibinfo{journal}{\emph{Journal of machine learning research}}
  \bibinfo{volume}{12}, \bibinfo{number}{Oct} (\bibinfo{year}{2011}),
  \bibinfo{pages}{2825--2830}.
\newblock


\bibitem[\protect\citeauthoryear{Porras, Sa\"{\i}di, and Yegneswaran}{Porras
  et~al\mbox{.}}{2009}]%
        {conficker2009}
\bibfield{author}{\bibinfo{person}{Phillip Porras}, \bibinfo{person}{Hassen
  Sa\"{\i}di}, {and} \bibinfo{person}{Vinod Yegneswaran}.}
  \bibinfo{year}{2009}\natexlab{}.
\newblock \showarticletitle{{A Foray into Conficker's Logic and Rendezvous
  Points}}. In \bibinfo{booktitle}{\emph{Proceedings of the 2Nd USENIX
  Conference on Large-scale Exploits and Emergent Threats: Botnets, Spyware,
  Worms, and More}} \emph{(\bibinfo{series}{LEET'09})}.
  \bibinfo{publisher}{USENIX Association}, \bibinfo{address}{Berkeley, CA,
  USA}, \bibinfo{pages}{7--7}.
\newblock
\urldef\tempurl%
\url{http://dl.acm.org/citation.cfm?id=1855676.1855683}
\showURL{%
\tempurl}


\bibitem[\protect\citeauthoryear{Schiavoni, Maggi, Cavallaro, and
  Zanero}{Schiavoni et~al\mbox{.}}{2014}]%
        {phoenix14}
\bibfield{author}{\bibinfo{person}{Stefano Schiavoni},
  \bibinfo{person}{Federico Maggi}, \bibinfo{person}{Lorenzo Cavallaro}, {and}
  \bibinfo{person}{Stefano Zanero}.} \bibinfo{year}{2014}\natexlab{}.
\newblock \showarticletitle{{Phoenix: DGA-Based Botnet Tracking and
  Intelligence}}. In \bibinfo{booktitle}{\emph{Detection of Intrusions and
  Malware, and Vulnerability Assessment}},
  \bibfield{editor}{\bibinfo{person}{Sven Dietrich}} (Ed.).
  \bibinfo{publisher}{Springer International Publishing},
  \bibinfo{address}{Cham}, \bibinfo{pages}{192--211}.
\newblock
\showISBNx{978-3-319-08509-8}


\bibitem[\protect\citeauthoryear{Shannon}{Shannon}{1948}]%
        {shannon48}
\bibfield{author}{\bibinfo{person}{C.~E. Shannon}.}
  \bibinfo{year}{1948}\natexlab{}.
\newblock \showarticletitle{A Mathematical Theory of Communication}.
\newblock \bibinfo{journal}{\emph{Bell Systems Technical Journal}}
  \bibinfo{volume}{27} (\bibinfo{year}{1948}), \bibinfo{pages}{623--656}.
\newblock


\bibitem[\protect\citeauthoryear{Shibahara, Yagi, Akiyama, Chiba, and
  Yada}{Shibahara et~al\mbox{.}}{2016}]%
        {Shibahara2016EfficientDM}
\bibfield{author}{\bibinfo{person}{Toshiki Shibahara}, \bibinfo{person}{Takeshi
  Yagi}, \bibinfo{person}{Mitsuaki Akiyama}, \bibinfo{person}{Daiki Chiba},
  {and} \bibinfo{person}{Takeshi Yada}.} \bibinfo{year}{2016}\natexlab{}.
\newblock \showarticletitle{Efficient Dynamic Malware Analysis Based on Network
  Behavior Using Deep Learning}.
\newblock \bibinfo{journal}{\emph{2016 IEEE Global Communications Conference
  (GLOBECOM)}} (\bibinfo{year}{2016}), \bibinfo{pages}{1--7}.
\newblock


\bibitem[\protect\citeauthoryear{Skuratovich}{Skuratovich}{2015}]%
        {skuratovich2015}
\bibfield{author}{\bibinfo{person}{Stanislav Skuratovich}.}
  \bibinfo{year}{2015}\natexlab{}.
\newblock \bibinfo{booktitle}{\emph{{MATSNU}}}.
\newblock \bibinfo{type}{{T}echnical {R}eport}. \bibinfo{institution}{Check
  Point Software Technologies Ltd.}
\newblock


\bibitem[\protect\citeauthoryear{Sp\"{a}rck~Jones}{Sp\"{a}rck~Jones}{1972}]%
        {sparck1972statistical}
\bibfield{author}{\bibinfo{person}{Karen Sp\"{a}rck~Jones}.}
  \bibinfo{year}{1972}\natexlab{}.
\newblock \showarticletitle{A statistical interpretation of term specificity
  and its application in retrieval}.
\newblock \bibinfo{journal}{\emph{Journal of documentation}}
  \bibinfo{volume}{28}, \bibinfo{number}{1} (\bibinfo{year}{1972}),
  \bibinfo{pages}{11--21}.
\newblock


\bibitem[\protect\citeauthoryear{Szegedy, Zaremba, Sutskever, Bruna, Erhan,
  Goodfellow, and Fergus}{Szegedy et~al\mbox{.}}{2013}]%
        {szegedy2013intriguing}
\bibfield{author}{\bibinfo{person}{Christian Szegedy},
  \bibinfo{person}{Wojciech Zaremba}, \bibinfo{person}{Ilya Sutskever},
  \bibinfo{person}{Joan Bruna}, \bibinfo{person}{Dumitru Erhan},
  \bibinfo{person}{Ian Goodfellow}, {and} \bibinfo{person}{Rob Fergus}.}
  \bibinfo{year}{2013}\natexlab{}.
\newblock \showarticletitle{Intriguing properties of neural networks}.
\newblock \bibinfo{journal}{\emph{arXiv preprint arXiv:1312.6199}}
  (\bibinfo{year}{2013}).
\newblock


\bibitem[\protect\citeauthoryear{Tong and Nguyen}{Tong and Nguyen}{2016}]%
        {semanticdga16}
\bibfield{author}{\bibinfo{person}{Van Tong} {and} \bibinfo{person}{Giang
  Nguyen}.} \bibinfo{year}{2016}\natexlab{}.
\newblock \showarticletitle{A method for detecting {DGA} botnet based on
  semantic and cluster analysis}. In \bibinfo{booktitle}{\emph{Proceedings of
  the Seventh Symposium on Information and Communication Technology, SoICT
  2016, Ho Chi Minh City, Vietnam, December 8-9, 2016}}.
  \bibinfo{pages}{272--277}.
\newblock
\urldef\tempurl%
\url{https://doi.org/10.1145/3011077.3011112}
\showDOI{\tempurl}


\bibitem[\protect\citeauthoryear{Woodbridge, Anderson, Ahuja, and
  Grant}{Woodbridge et~al\mbox{.}}{2016}]%
        {woodbridge2016predicting}
\bibfield{author}{\bibinfo{person}{Jonathan Woodbridge},
  \bibinfo{person}{Hyrum~S Anderson}, \bibinfo{person}{Anjum Ahuja}, {and}
  \bibinfo{person}{Daniel Grant}.} \bibinfo{year}{2016}\natexlab{}.
\newblock \showarticletitle{Predicting domain generation algorithms with long
  short-term memory networks}.
\newblock \bibinfo{journal}{\emph{arXiv preprint arXiv:1611.00791}}
  (\bibinfo{year}{2016}).
\newblock


\bibitem[\protect\citeauthoryear{Yadav, Kumar, Reddy, L.~Narasimha~Reddy, and
  Ranjan}{Yadav et~al\mbox{.}}{2012}]%
        {domainflux12}
\bibfield{author}{\bibinfo{person}{Sandeep Yadav}, \bibinfo{person}{Ashwath
  Kumar}, \bibinfo{person}{Krishna Reddy}, \bibinfo{person}{A
  L.~Narasimha~Reddy}, {and} \bibinfo{person}{Supranamaya Ranjan}.}
  \bibinfo{year}{2012}\natexlab{}.
\newblock \showarticletitle{{Detecting Algorithmically Generated Domain-Flux
  Attacks With DNS Traffic Analysis}}.
\newblock   \bibinfo{volume}{20} (\bibinfo{date}{10} \bibinfo{year}{2012}).
\newblock


\bibitem[\protect\citeauthoryear{Yu, Pan, Hu, Nascimento, and De~Cock}{Yu
  et~al\mbox{.}}{2018}]%
        {yu2018character}
\bibfield{author}{\bibinfo{person}{Bin Yu}, \bibinfo{person}{Jie Pan},
  \bibinfo{person}{Jiaming Hu}, \bibinfo{person}{Anderson Nascimento}, {and}
  \bibinfo{person}{Martine De~Cock}.} \bibinfo{year}{2018}\natexlab{}.
\newblock \showarticletitle{Character Level based Detection of {DGA} Domain
  Names}. In \bibinfo{booktitle}{\emph{Proceedings of the 2018 International
  Joint Conference on Neural Networks (IJCNN '18)}}.
\newblock


\bibitem[\protect\citeauthoryear{Zhao, Xu, Xu, and Wu}{Zhao
  et~al\mbox{.}}{2015}]%
        {aptdns15}
\bibfield{author}{\bibinfo{person}{Guodong Zhao}, \bibinfo{person}{Ke Xu},
  \bibinfo{person}{Lei Xu}, {and} \bibinfo{person}{Bo Wu}.}
  \bibinfo{year}{2015}\natexlab{}.
\newblock \showarticletitle{{Detecting APT Malware Infections Based on
  Malicious DNS and Traffic Analysis}}.
\newblock   \bibinfo{volume}{3} (\bibinfo{date}{01} \bibinfo{year}{2015}),
  \bibinfo{pages}{1132--1142}.
\newblock


\bibitem[\protect\citeauthoryear{Zhou, Li, Miao, and Yim}{Zhou
  et~al\mbox{.}}{2013}]%
        {Zhou2013DGABasedBD}
\bibfield{author}{\bibinfo{person}{Yonglin Zhou}, \bibinfo{person}{Qing-Shan
  Li}, \bibinfo{person}{Qidi Miao}, {and} \bibinfo{person}{Kangbin Yim}.}
  \bibinfo{year}{2013}\natexlab{}.
\newblock \showarticletitle{{DGA-Based Botnet Detection Using DNS Traffic}}.
\newblock \bibinfo{journal}{\emph{J. Internet Serv. Inf. Secur.}}
  \bibinfo{volume}{3} (\bibinfo{year}{2013}), \bibinfo{pages}{116--123}.
\newblock


\end{thebibliography}

\end{document}